\documentclass[]{aastex63}
\usepackage{graphicx}
\usepackage{subfigure}
\usepackage{multirow}
\usepackage{mathptmx} 
\usepackage{hyperref}

\let\counterwithin\relax
\usepackage{chngcntr} 

\hypersetup{linkcolor=blue,citecolor=blue,filecolor=cyan,urlcolor=blue}

\newcommand{\MyTitle}{The DR21(OH) trident --- resolving the massive ridge into three entangled fibers as the initial condition of cluster formation}

\newcommand{\NJU}{School of Astronomy and Space Science, Nanjing University, 163 Xianlin Avenue, Nanjing 210023, People's Republic of China}
\newcommand{\NJULab}{Key Laboratory of Modern Astronomy and Astrophysics (Nanjing University), Ministry of Education, Nanjing 210023, People's Republic of China}
\newcommand{\CfA}{Center for Astrophysics $\vert$ Harvard \& Smithsonian, 60 Garden Street, MS 42, Cambridge, MA 02138, USA}
\newcommand{\YNU}{South-Western Institute for Astronomy Research, Yunnan University, Kunming, 650500 Yunnan, People's Republic of China}

\newcommand{\hii}{H{\scriptsize\ II}}
  
\newcommand{\degree}{$^\circ$} 
\newcommand{\msun}{$M_{\odot}$}
\newcommand{\msunpc}{\msun\ $\rm pc^{-1}$}

\newcommand{\um}{$\rm{\mu m}$}
\newcommand{\kms}{$\rm km\ s^{-1}$} 
\newcommand{\kmspc}{$\rm km\ s^{-1}pc^{-1}$} 
\newcommand{\mjybeam}{$\rm mJy\ beam^{-1}$}

\newcommand{\moleA}{$\rm H^{13}CO^+$}
\newcommand{\moleB}{$\rm N_2H^+$}
\newcommand{\moleC}{$\rm NH_2D$}
\newcommand{\lineA}{\moleA\ ($1-0$)}
\newcommand{\lineB}{\moleB\ ($1-0$)}
\newcommand{\lineC}{\moleC\ ($1_{1,1}-1_{0,1}$)}

\newcommand{\fa}{f1} 
\newcommand{\fb}{f2} 
\newcommand{\fc}{f3} 

\newcommand{\gradvMoleA}{1.0 \kmspc} 
\newcommand{\xStretchMoleA}{2.8 pc} 
\newcommand{\gradvMoleB}{0.8 \kmspc} 
\newcommand{\xStretchMoleB}{0.5 pc} 
\newcommand{\gradvMoleC}{2.0 \kmspc} 
\newcommand{\xStretchMoleC}{1.0 pc} 

\newcommand{\numSpectraMoleA}{7,625}
\newcommand{\numSpectraMoleB}{7,500}
\newcommand{\numSpectraMoleC}{2,141}

\newcommand{\numPointMoleA}{10,152}
\newcommand{\numPointMoleB}{7,766}
\newcommand{\numPointMoleC}{2,141}

\newcommand{\beamHerschel}{12\arcsec.6} 
\newcommand{\beamJCMTA}{7\arcsec.9} 
\newcommand{\beamJCMTB}{13\arcsec} 
\newcommand{\beamColden}{13\arcsec.7} 
\newcommand{\beamColdenSix}{38\arcsec.5} 

\newcommand{\massFa}{5117} 
\newcommand{\massFb}{1838} 
\newcommand{\massFc}{921} 
\newcommand{\lineMassFa}{2133} 
\newcommand{\lineMassFb}{795} 
\newcommand{\lineMassFc}{511} 

\newcommand{\Mp}{582.1} 
\newcommand{\MpErr}{13.9} 
\newcommand{\vp}{-3.87} 
\newcommand{\vpErr}{0.04}

\submitjournal{ApJ}

\begin{document} 

\title{\MyTitle} 

\author[0000-0002-6368-7570]{Yue Cao}
\affiliation{\NJU}\affiliation{\NJULab}\affiliation{\CfA}
\author[0000-0002-5093-5088]{Keping Qiu}
\affiliation{\NJU}\affiliation{\NJULab}
\author[0000-0003-2384-6589]{Qizhou Zhang}
\affiliation{\CfA}
\author[0000-0003-3144-1952]{Guang-Xing Li}
\affiliation{\YNU}

\correspondingauthor{Keping Qiu}
\email{kpqiu@nju.edu.cn} 


\begin{abstract} 
DR21(OH) ridge, the central part of a high-mass star and cluster forming hub-filament system, is resolved spatially and kinematically into three nearly parallel fibers (\fa, \fb, and \fc) with a roughly north-south orientation, using the observations of molecular transitions of \lineA, \lineB, and \lineC\ with the Combined Array for Research in Millimeter Astronomy. These fibers are all mildly supersonic ($\sigma_{\rm V}$ about 2 times the sound speed), having lengths around 2~pc and widths about 0.1~pc, and they entangle and conjoin in the south where the most active high-mass star formation takes place. They all have line masses 1--2 orders of magnitude higher than their low-mass counterparts and are gravitationally unstable both radially and axially. However, only \fa\ exhibits high-mass star formation all the way along the fiber, yet \fb\ and \fc\ show no signs of significant star formation in their northern parts. A large velocity gradient increasing from north to south is seen in \fc, and can be well reproduced with a model of free-fall motion toward the most massive and active dense core in the region, which corroborates the global collapse of the ridge and suggests that the disruptive effects of the tidal forces may explain the inefficiency of star formation in f2 and f3. On larger scales, some of the lower-density, peripheral filaments are likely to be the outer extensions of the fibers, and provide hints on the origin of the ridge.
\end{abstract}


\section{Introduction}\label{sec:intro} 

Filamentary structures have been widely known to pervade the cold interstellar medium \citep[e.g.,][]{2010A&A...518L.102A,2014prpl.conf...27A,2010A&A...518L.100M,2014ApJ...791...27S}. While molecular filaments span a wide range in length and density, over the past decade dense filaments of clump to cloud scales, i.e., 1--10~pc, are extensively studied as they may bridge relatively diffuse molecular gas and compact cores capable of forming individual stars or multiple stars. Although it is not impossible to find filaments in relative isolation \citep[e.g.,][]{2016A&A...587A..97H}, more often they appear to be organized into more complex structures, such as a web-like network or nest \citep[e.g.,][]{2010A&A...518L.103M,2011A&A...529L...6A,2011A&A...533A..94H,2013ApJ...764L..26B}, or a system of several filaments converging at a high-density region, i.e., a hub-filament system \citep[HFS, e.g.,][]{2010ApJ...725...17G,2012ApJ...756...10L,2012A&A...540L..11S,2013ApJ...766..115K,2014A&A...561A..83P,2018A&A...613A..11W}. Massive and elongated hub regions are sometimes referred to as ``ridges'' \citep[e.g.,][]{2011A&A...533A..94H,2012A&A...543L...3H,2017A&A...602A..77T,2018ARA&A..56...41M}. HFSs are of particular interests to the formation of massive stars and clusters since the hubs or ridges are located close to the gravity center and thus may aid in global collapse or large-scale gas inflow along the filaments \citep{2009ApJ...700.1609M,2014prpl.conf...27A,2020A&A...642A..87K}. And observed velocity fields in some massive filaments and HFSs are consistent with global collapse at a high accretion rate of order $10^{-3}$~\msun\,${\rm yr}^{-1}$  \citep[e.g.,][]{2010A&A...520A..49S,2013A&A...554L...2Z,2021ApJ...908...70H}, pointing to a dynamic picture of cluster formation \citep{2019ARA&A..57..227K}. Relatively diffuse molecular gas also show filamentary features, and such low-density, tenuous ``striations'' were first observed in CO lines \citep{2008ApJ...680..428G}, and then seen in dust continuum with \emph{Herschel} observations \citep{2013A&A...550A..38P}. 

There is growing evidence that filaments are not monolithic but have substructures. \citet{2013A&A...554A..55H} identified multiple velocity components in position-position-velocity (PPV) space in the gas of the low-mass star-forming filament in Taurus, and suggested that the filament is indeed a bundle of those ``velocity-coherent structures'', which are also termed ``fibers''. After that work, similar fiber-like substructures have been detected in other dense, both low-mass and high-mass, molecular filaments \citep{2014MNRAS.440.2860H,2016A&A...590A..75F,2018A&A...610A..77H,2019A&A...632A..83S}, while the physical origin of the fibers remains unclear \citep{2015A&A...574A.104T,2017MNRAS.468.2489C}. The methodologies used to extract fibers (or fiber-like substructures) all work in PPV space, but vary in practice for the detailed techniques. There is caution that features identified in PPV space do not really represent density structures in position-position-position (PPP) space \citep[see][]{2018MNRAS.479.1722C}, but this issue could be mitigated with interferometer high angular resolution observations which have started to spatially resolve the internal structure of a filament. Filament substructures may also play a role in the formation and dynamical evolution of hubs or ridges in massive HFSs. \citet{2017MNRAS.464L..31H} observed a protostellar hub within an massive infrared dark cloud with the Atacama Large Millimeter/submillimeter Array (ALMA) in the 1~mm continuum, and detected multiple intra-hub subfilaments, which are narrow (0.028~pc) and analogous to fibers. \citet{2018A&A...610A..77H} identified dozens of dense fibers in the central region of the well-known Integral-Shape Filament (ISF) in Orion, and recognized multiple hub-like associations along a dense bundle of the fibers. Nevertheless, observational studies capable of resolving massive filaments and hubs are still very limited, leaving several key questions in the context unsolved: (1) what are the physical properties of the substructures in massive HFSs and how do they differ from their low-mass counterparts? (2) What a role do these substructures play in shaping the initial conditions of high-mass star and cluster formation? (3) How are they related to large-scale filaments connected to the peripherals of the hubs or ridges? To shed light on these questions, we carry out interferometer molecular spectral line observations toward the central ridge of a pronounced massive HFS, and successfully resolve its inner structures both spatially and kinematically. By deriving the physical properties of the substructures and investigating their kinematics and dynamics, we try to understand their relation with the ongoing high-mass star formation.

The DR21 ridge (also called DR21 filament) is an elongated massive gas structure connected by several filaments with varying orientations, forming a remarkable HFS in the Cygnus X molecular cloud complex \citep{2007A&A...476.1243M,2010A&A...520A..49S,2012A&A...543L...3H}. It contains a massive clump in the southern end and a filamentary structure to the north. The southern clump is embedded with at least two compact \hii\ regions, namely DR21 \citep{2003ApJ...596..344C}. The filamentary structure in the north contains a chain of $\sim$0.1~pc massive dense cores (MDCs), and the most massive one is DR21(OH), which is associated with various masers and actively forming a cluster of high-mass stars \citep{2012ApJ...744...86Z}. While the DR21 ridge in the literature refers to the whole structure including both DR21 and DR21(OH), in this work we only focus on the continuous structure in the north; we leave out the DR21 clump for its discontinuity with the northern part and its more advanced evolutionary stage (thus of less relevance to the initial conditions of high-mass star formation). And to avoid confusion, we refer to this structure as the ``DR21(OH) ridge''. Located at a distance of 1.5 kpc \citep{2012A&A...539A..79R}, this structure has a projected length of $\sim$2.5~pc and a total gas mass of $\sim$9000 \msun, making its line mass two orders of magnitude higher than those of the low-mass filaments \citep[e.g.][]{2011A&A...533A..94H,2013A&A...554A..55H} and about 10 times higher than that of the ISF in Orion \citep{2021A&A...651A..36S}. Given its relatively close distance and great mass, the DR21(OH) ridge is an ideal target for disentangling the substructures of high-mass cluster-forming HFSs.

\section{Observations} \label{sec:obs}

The DR21(OH) ridge was observed in the 3~mm waveband, covering \lineA, \lineB, and \lineC, with the 15-antenna Combined Array for Research in Millimeter-wave Astronomy (CARMA) in the D configuration during June 29--July 14, 2014 (Project ID: c1212; PI: Keping Qiu). The CARMA D array provides a baseline range of 11--148 m, yielding a synthesized beam size of $\sim5$\arcsec\ (0.036 pc@1.5 kpc) and a largest recovering scale of $\sim$70\arcsec\ (0.5 pc@1.5 kpc). Mosaic observations toward four phase centers (R.A.=$20^{\rm h}39^{\rm m}01^{\rm s}.0$; Decl.=42\degree22\arcmin25\arcsec.5, 42\degree23\arcmin25\arcsec.5, 42\degree24\arcmin25\arcsec.5, and 42\degree25\arcmin25\arcsec.5) were conducted to map the ridge. Observations toward each phase center were made with one or two scheduling blocks, leading to a total observing time of about 5 to 6 hours and an on-source integration time of 2 to 3 hours each. Spectral windows \#7, \#4, and \#8 of the CARMA correlator were centered at the rest frequencies of the \moleA, \moleB, and \moleC\ transitions, respectively, with a uniform bandwidth of 31.152 MHz and spectral resolution of 97.7 kHz (0.33 \kms@90 GHz; 319 channels). J1927+739 was used as the bandpass calibrator and MWC349 was used for both flux calibration and time-dependent phase and amplitude gain calibration. The raw data were flagged, calibrated, and imaged with the Miriad software \citep{1995ASPC...77..433S}. We used the Briggs weighting with the robust parameter set to 0.5 to image the data. The final data products contain three mosaic spectral line cubes with a map size of about 3\arcmin$\times$6\arcmin\ and a pixel size of 1\arcsec.5. The velocity resolutions are 0.34, 0.31, and 0.34 \kms, and the 1-$\sigma$ noise levels are 36.9, 42.4, and 34.7 \mjybeam\ per velocity channel for the \moleA, \moleB, and \moleC\ data cubes, respectively. Detailed observational setups and logs can be found at the CARMA Data Archive\footnote{\url{http://carma-server.ncsa.uiuc.edu:8181/}}.

\section{Results, Analysis, and Discussion}\label{sec:analysis}

Figure~\ref{fig:channel} shows the velocity channel maps of the \lineA\ emission, where the original channel width of 0.34~\kms is smoothed to 1.0~\kms. The \moleA emission is detected at $-6$ to 1~\kms, with the emission at $\lesssim-3.0$~\kms mostly traces the high density peaks as visualized by a chain of active MDCs \citep{2007A&A...476.1243M}. On the other hand, the emission at $\gtrsim-2.0$~\kms reveals new features to the west of the MDC chain. Overall the results here are consistent with the single-dish \lineA\ observations by \citet{2010A&A...520A..49S}, except that our data have a factor of 6 better angular resolution. Because of the missing zero-spacing information, our interferometer observations filter out extended emission (larger than 0.5 pc, see Section \ref{sec:obs}). We anticipate, however, that the extraction and analysis of the substructures within the DR21(OH) ridge are not strongly affected. We do not show the velocity channel maps in \lineB\ and \lineC, since these transitions have blended hyperfine lines. The velocity-integrated emissions of the three lines are all highly biased to the structure coincident with the MDC chain (Figure~\ref{fig:m0}). Thus, more sophisticated analyses of the spectral line data are needed to better unraveling the velocity structure of the ridge (see Section \ref{subsec:fit}).

\subsection{High-resolution $\rm H_2$ Column Density Map and Dust Temperature Map Derived from Continuum Data} \label{subsec:N_map} 

\begin{figure*}[htb!]
\epsscale{1.2}\plotone{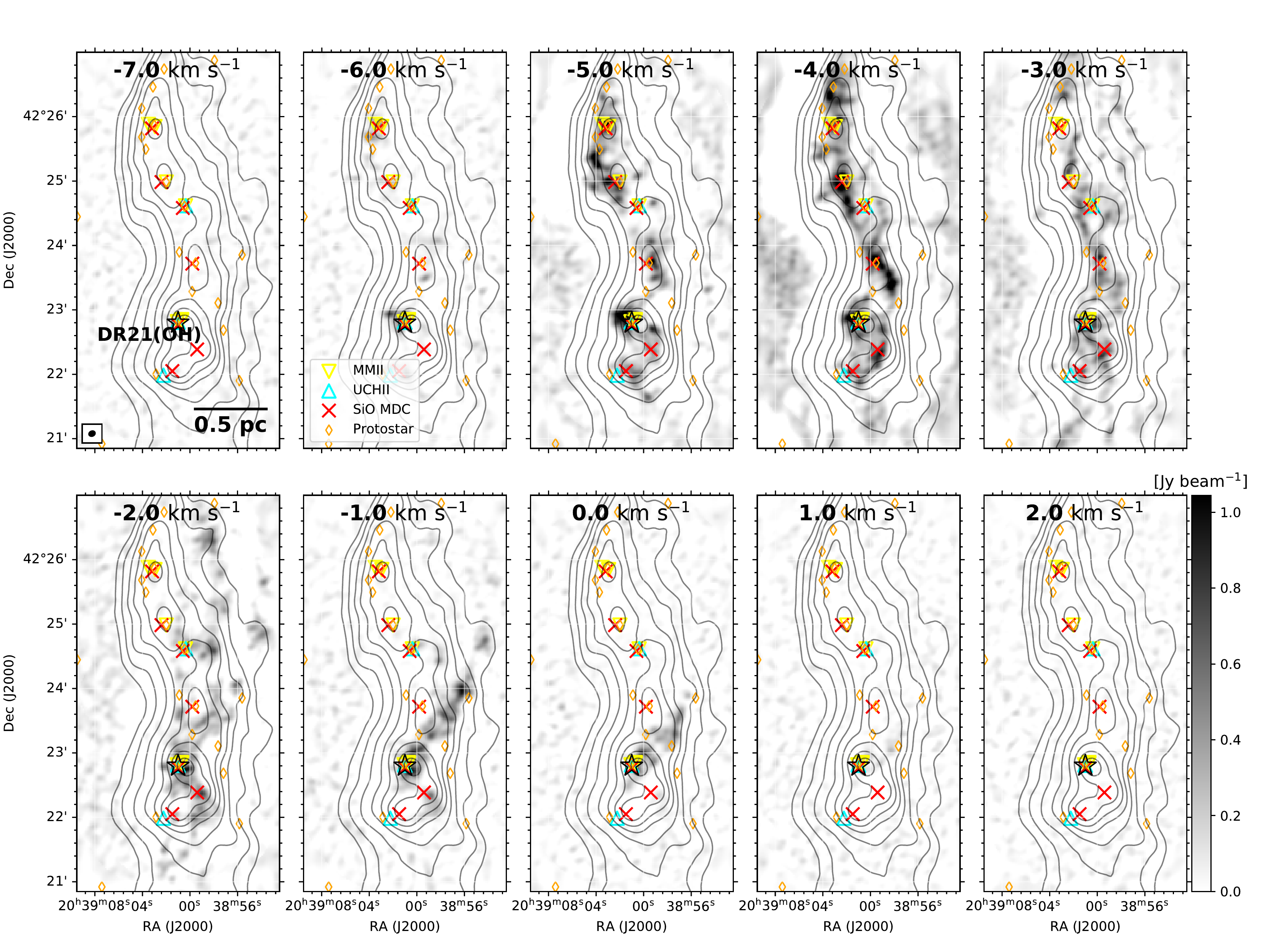}
\caption{Channel maps of the \lineA\ emission shown in gray scale, and the scaling is shown in a vertical bar in the lower-right corner. The central velocity of each channel is labeled on the top of each panel. Black contours of $N_{\rm H_2}$ in each panel derived with submillimeter continuum data (Sect. \ref{subsec:N_map}) are drawn at $1.5^{[0,1,2,3,...]}\times5\times10^{22} \rm cm^{-2}$. In all the panels, catalogs of class II methanol masers (MMII) and ultracompact \hii\ (UC~\hii) regions are from \citet{2019ApJS..241....1C}, and are marked as yellow inverted triangles and cyan triangles, respectively; massive dense cores with strong SiO emissions tabulated by \citet{2007A&A...476.1243M} are plotted as red crosses; protostars in \citet{2007MNRAS.374...29D} and \citet{2014AJ....148...11K} are marked as orange diamonds. The location of DR21(OH) is marked as a black star. A linear size scale bar and the CARMA beam are shown in the most upper left panel.}\label{fig:channel}
\end{figure*}

To calculate the mass and other physical properties, high-resolution (\beamColden) $\rm H_2$ column density ($N_{\rm H_2}$) and dust temperature ($T_{\rm dust}$) maps of the DR21(OH) ridge are derived by fitting the dust continuum emissions at submillimeter wavelengths with a modified blackbody model \citep{1983QJRAS..24..267H}:

\begin{equation}\label{eq:dust_I}
I_{\nu}=B_{\nu}(T_{\rm dust})(1-e^{-\tau_{\nu}}),
\end{equation}

\begin{equation}\label{eq:dust_tau}
\tau_{\nu}=\mu_{\rm H_2}m_{\rm H}N_{\rm H_2}\kappa_{\nu}/\Gamma,
\end{equation}

\noindent where $B_{\nu}(T_{\rm dust})$ is the Planck function, $\mu_{\rm H_2}=2.8$ is the mean molecular weight per $\rm H_2$ molecule, $m_{\rm H}$ is the mass of the hydrogen atom, $\kappa_{\nu}$ is the dust mass opacity, and $\Gamma=100$ is a canonical gas-to-dust mass ratio for the interstellar medium. We evaluate $\kappa_{\nu}$ following $\kappa_{\nu}=\kappa_0 (\nu/\nu_0)^{\beta}$, where $\kappa_0=10\rm\ cm^2g^{-1}$, $\nu_0=1\rm\ THz$, and $\beta=2$. This dust opacity law is widely used by the \emph{Herschel} large survey projects \citep[e.g.,][]{2010A&A...518L.102A,2017A&A...602A..77T}. The continuum maps used for the fitting were obtained from \citet{2019ApJS..241....1C} which include the \emph{Herschel}/PACS 160 \um\ map and the James Clerk Maxwell Telescope (JCMT)/SCUBA-2 450 and 850 \um\ maps. The beam sizes of the 160, 450, and 850 \um\ maps are \beamHerschel, \beamJCMTA, and \beamJCMTB, respectively. We did not use the \emph{Herschel} 70 \um\ data due to the possible contaminations from the emissions of very small grains \citep{2001ApJ...551..807D} that are not considered in the single-temperature model \citep[see Figure 12 of ][]{2019ApJS..241....1C}. The \emph{Herschel} maps at 250, 350, and 500 \um\ were not used since their resolutions are too coarse ($>18$\arcsec) compared with the CARMA data. The continuum maps used for the fitting were cropped to the same size as the CARMA data (3\arcmin$\times$6\arcmin), smoothed to a common beam size of \beamColden, and resampled to the same griding with a pixel size of 2\arcsec. The flux uncertainties of the continuum maps are conservatively estimated to be 20\% following \citet{2019ApJS..241....1C} and are considered in the fitting procedure. We use the \emph{minimize} function in the Scipy\footnote{\url{https://www.scipy.org/}} \citep{scipy} package to implement the fitting. The derived $N_{\rm H_2}$ map is shown in Figures~\ref{fig:channel} and \ref{fig:m0}, and the $T_{\rm dust}$ map is shown in Figure~\ref{fig:m0} (also see the $N_{\rm H_2}$ and $T_{\rm dust}$ uncertainties in Appendix \ref{app:6band}). To test the robustness of the results derived from the 3-band fitting, we generate another set of $N_{\rm H_2}$ and $T_{\rm dust}$ maps with 6-band data (\emph{Herschel} 160, 250, 350, and 500 \um; JCMT 450 and 850 \um) and compare them at the same resolution (see Appendix \ref{app:6band}). We find that 93\% of the total map area has $T_{\rm dust}$ differences less than 2 K, and 90\% of the map area has relative differences in $N_{\rm H_2}$ less than 25\%, indicating that the 3-band fitting is robust. 

$N_{\rm H_2}$ and $T_{\rm dust}$ in the DR21(OH) ridge range from $5\times10^{21}$ to $1\times10^{24}\rm\ cm^{-2}$ and from 15 to 31 K, respectively, where the highest column density and temperature are both found toward the DR21(OH) core. The ridge can be defined with a $N_{\rm H_2}$ contour level of $10^{23}\rm\ cm^{-2}$ as suggested by \citet{2012A&A...543L...3H}. In Figure~\ref{fig:channel}, if we define the outer edge of the ridge with a slightly lower $N_{\rm H_2}$ value of $5\times10^{22}\rm\ cm^{-2}$, the \moleA\ emissions, including the bright emission tracing the MDC chain and the new features to the west, are all confined to be within the ridge. In Figure~\ref{fig:m0}(a), to the west of the MDC chain, the 8~\um\ dark patches (absorption) coincident with the \moleA\ emission features are also perceptible. All this indicates that \moleA emission is probing high density substructures within the DR21(OH) ridge.

\subsection{Spectral Fitting and the PPV Structures of the DR21(OH) ridge} \label{subsec:fit} 

To exact the PPV structures and derive the physical properties of the DR21(OH) ridge, we fit the data cubes of the three transitions pixel-by-pixel with a multi-velocity-component spectral line model based on the theoretical work of \citet{2015PASP..127..266M}. In this model, the $i$th velocity component along the line-of-sight (LoS) of a pixel has a total column density $N_{{\rm trc},i}$ for a certain tracer and an excitation temperature $T_{\rm ex}$. From Eq. (32) of \citet{2015PASP..127..266M}, we have the velocity distribution of the tracer column density

\begin{equation}\label{eq:N_fun}
N_{{\rm trc},i}({\rm v}) = \frac{3h}{8\pi^3 S \mu^{2}} \frac{Q_{\rm rot}}{g_u} e^{\frac{E_u}{kT_{\rm ex}}}\left(e^{\frac{h\nu_{0i}}{k_{\rm B}T_{\rm ex}}}-1\right)^{-1}\tau_{\nu i}({\rm v}),
\end{equation}
where $h$ is the Planck constant, $k_{\mathrm{B}}$ is the Boltzmann constant, and $S$, $\mu$, $g_{u}$, $Q_{\mathrm{rot}}$, $E_{\mathrm{u}}$, $\nu_{0}$ are the line strength, dipole moment, degeneracy of the upper energy level, rotational partition function, upper level energy, and rest frequency of the transition, respectively. We assume in the model that $N_{{\rm trc},i}({\rm v})$ has a Gaussian distribution with a centroid velocity ${\rm v}_{0i}$ and a velocity dispersion $\sigma_{\rm vi}$, leading to

\begin{equation}\label{eq:N_gauss}
N_{{\rm trc},i}({\rm v}) = \frac{N_{{\rm trc},i}}{\sqrt{2\pi}\sigma_{\rm vi}} e^{-\frac{({\rm v}-{\rm v}_{0i})^2}{2\sigma_{\rm vi}^2}}.
\end{equation}

Further taking into account hyperfine lines for a transition, we derive the opacity of the $j$th hyperfine line of a transition for the $i$th velocity component

\begin{equation}\label{eq:tau}
\tau_{\nu i j}({\rm v})=\frac{4 \sqrt{2\pi^5} S \mu^{2} R_{j} g_{u}}{3 h Q_{\rm rot} \sigma_{{\rm v} i}} N_{{\rm trc},i} \left(e^{\frac{h \nu_{0}}{k_{\mathrm{B}} T_{\mathrm{ex}}}}-1\right) e^{-\frac{\left({\rm v}-{\rm v}_{0 i}-\delta {\rm v}_j\right)^{2}}{2 \sigma_{{\rm v} i}^{2}}-\frac{E_{\rm u}}{k_{\rm B} T_{\rm ex}}},
\end{equation}

\noindent where $R_{j}$ and $\delta{\rm v}_j$ are the relative line strength and the velocity offset relative to $\nu_{0}$ for the $j$th hyperfine line, respectively Here we assume that all the velocity components along the LoS of one pixel have a common $T_{\rm ex}$, otherwise the radiative transfer equation is not analytically integrable and the detailed material distribution along the LoS must be known. The modeled spectral intensity as a function of velocity is given by

\begin{equation}\label{eq:intensity}
\Delta I_{\nu}\left({\rm v}; N_{{\rm trc},i}, T_{\rm ex}, {\rm v}_{0i}, \sigma_{{\rm v}i}\right)=\left(B_{\nu}\left(T_{\rm ex}\right)-B_{\nu}\left(T_{\rm bg}\right)\right)\left(1-e^{-\sum_{i j} \tau_{\nu i j}(\rm v)}\right), 
\end{equation}

\noindent where $T_{\mathrm{bg}}$ is the background brightness temperature and is set to the cosmic microwave background value of 2.73 K. Note that this spectral model is quite general and does \emph{not} rely on additional assumptions such as the Rayleigh-Jeans approximation or the optically thin approximation. By fitting an observed spectrum with this model one can obtain a parameter set ($N_{{\rm trc},i}$, $T_{\rm ex}$, ${\rm v}_{0i}$, $\sigma_{{\rm v}i}$) for each velocity component toward a pixel. 

Since we only have observations of one transition for one molecular species, $T_{\rm ex}$ and $N_{\rm trc}$ cannot be determined simultaneously in the fitting due to degeneration. Therefore we fix $T_{\rm ex}$ to $T_{\rm dust}$ obtained in Sect. \ref{subsec:N_map} for each pixel during the fitting procedure, by assuming that the gas and dust temperatures are equal. This assumption is valid if the gas and dust are strongly coupled \citep[e.g.,][]{{1983ApJ...265..223B}}. At low to intermediate densities ($n\lesssim10^4$~cm$^{-3}$), relatively weak gas-dust coupling and the depletion of coolant species may conspire a considerably higher gas temperature than the dust temperature; such an effect can be significantly reduced at higher densities thanks to rapid gas-dust coupling, leading to a temperature difference of $\sim$4~K at $n\sim10^5$~cm$^{-3}$ and completely negligible at $n\sim10^6$~cm$^{-3}$ \citep{2001ApJ...557..736G}. Dust evolution in dense cores can affect thermal gas-dust coupling, leading to a higher gas temperature compared to the dust temperature, but the temperature difference is also small at high densities \citep[$\sim$3~K at $n\sim10^5$~cm$^{-3}$ and $\sim$1~K at $n\sim10^6$~cm$^{-3}$,][]{2019ApJ...884..176I}. In our case, we have gas densities $>10^5$~cm$^{-3}$ ($N_{\rm H_2}$ in Figure~\ref{fig:channel} divided by 0.1~pc, which is derived in Sect. \ref{subsec:char}, also see Table~\ref{tab:filament}). Thus the dust temperature approximates the gas temperature within a few K. We further assess in Appendix~\ref{app:mass} that a small variation of 4~K in $T_{\rm ex}$ induces a 12\% difference in the fitted $N_{{\rm trc},i}$. We use the tool \emph{curve\_fit} in Scipy to extract and fit \numSpectraMoleA, \numSpectraMoleB, and \numSpectraMoleC\ spectra and obtain a total of \numPointMoleA, \numPointMoleB, and \numPointMoleC\ velocity components for the transitions of \moleA, \moleB, and \moleC, respectively (note that one pixel can have multiple velocity components). These velocity components can be identified as points in the PPV space (hereafter ``PPV points''). Figures~\ref{fig:ppv} provides 2D views of the PPV points. While projected on the plane of sky (PoS), these PPV points are grouped into several fiber-like substructures which entangle around the DR21(OH) core and branches to the north (Figures~\ref{fig:ppv}(a--c)). In the position-velocity (PV) diagrams (Figures~\ref{fig:ppv}(d--f)), the PPV points exhibit complicated velocity structures, which might be a consequence of supersonic turbulence. A particular interesting feature seen in the \moleA\ emission is the westernmost ``branch'' in Figure~\ref{fig:ppv}(a); we will come back to this substructure later by modeling its PV structure as revealed in Figure~\ref{fig:ppv}(d).

\subsection{Identification of the Fibers} \label{subsec:iden}

To identify the fibers seen in the PPV space, we apply the agglomerative clustering implementation in the scikit-learn package\footnote{\url{https://scikit-learn.org/stable/}} \citep{scikit-learn} to the PPV points for each transition. The agglomerative clustering algorithm groups points in N-dimensional space via recursively merging points or clusters (i.e. groups of points) into higher-order clusters such that the pair of points or clusters to be merged minimally increases the linkage distance \citep{wardclustering}. To make the clustering procedure adjustable, we introduce two parameters $t_x$ and $t_v$ in defining the linkage distance of two PPV points:

\begin{equation}\label{eq:distCluster}
s_{{\rm\ link},ij}=\sqrt{[t_x(x_i-x_j)]^2+(y_i-y_j)^2+\left(\frac{v_i-v_j}{t_v}\right)^2},
\end{equation}

\noindent where $x$ and $y$ are the spatial coordinates in \emph{physical} units, and ${\rm v}$ is the LoS velocity. For each transition, we adjust the parameters ($t_x$, $t_v$) and run the algorithm with the modified coordinates of the PPV points as input. By practice we found that the best values of ($t_x$, $t_v$) for identifying the fiber structures in the \moleA, \moleB, and \moleC\ data are (\xStretchMoleA, \gradvMoleA), (\xStretchMoleB, \gradvMoleB), and (\xStretchMoleC, \gradvMoleC), respectively, with which the DR21(OH) ridge is decomposed into 3, 3, and 2 fibers in the PPV space. Figure \ref{fig:rgb} shows the identified fibers projected on the PoS and their PV plots, with the color coding representing different velocities. In the \moleA\ data, three fibers with roughly north-south orientations are clearly seen; these fibers have distinct velocities, forming a ``trident'' with the junction approximately coincident with the DR21(OH) core. We name the three fibers as \fa, \fb, and \fc\ from east to west. Among the fibers, \fa\ is clearly tracing the central densest part of the dust ridge well known from previous dust continuum observations \citep{2007A&A...476.1243M,2012A&A...543L...3H}; f2 and f3 are relatively new, not seen in the dust emission, but previous single-dish \lineA\ observations showed that the ridge slight moves from east to west with the velocity increasing from $-5$~\kms to $0$~\kms \citep{2010A&A...520A..49S}, in a manner consistent with the positions and velocities of the three fibers identified here. In particular, f3 is also discernible in the \lineA\ map in \citet{2010A&A...520A..49S} (see their Figure 9), though at a much lower resolution. Given the LoS velocities of the fibers and the existing observations suggesting that the DR21(OH) ridge is in global collapse \citep{2010A&A...520A..49S}, one may expect that \fa\ and \fc\ are on the far side and the near side along LoS, respectively, and that \fb\ is probably in the middle. The three fibers in the \moleB\ line are less prominent yet are still clearly seen with positions and velocities consistent with those in the \moleA\ line, indicating that they are tracing the same physical entities. While the whole \fa\ and the southern part of \fc\ are detected, \fb\ is fragmented into several parts, probably due to the regional variations of the abundances of the two molecular tracers. On the other hand, in the \moleC\ line, the whole \fa\ (though fragmented into to parts) and the northern part of \fb\ are detected, and \fc\ is completely absent. The appearance of the fibers are also different from that in the \moleA\ and \moleB\ lines by their more compact morphologies and smaller widths. In addition, in contrast to what is seen in the \moleA\ and \moleB\ lines, the southern end of \fa\ in the \moleC\ line is shifted to the west and does not coincide with the DR21(OH) core, indicating that active high-mass star formation has destroyed most of \moleC. The detection information of the fibers in the three transitions is summarized in Table \ref{tab:filament}.

\subsection{Characterizing the Properties of the Fibers}\label{subsec:char}

In this subsection we derive the physical properties of the fibers with the \moleA\ data due to their best detections. We first determine the major axes of the fibers in the PoS through linear regression analyses of the coordinates of the PPV points, and divided the PPV points of each fiber into 15 equal-length parts along the major axis. For each part a ``bone'' point is generated with flux-weighted mean positions and other physical parameters ($N_{\rm trc}$, $T_{\rm ex}$, ${\rm v}_0$, $\sigma_{\rm v}$) of the PPV points in that part. The derived bone points of the three fibers are shown in Figure \ref{fig:rgb}, which delineate the axes of the fibers in the PPV space and can be used for calculating the distribution of the physical properties along the fibers. Lengths of the fibers are estimated as the lengths of the bone lines, and the widths are evaluated as the dispersions of the distances of the PPV points to the bone lines. Other physical quantities (e.g. the mean LoS velocity and velocity dispersion) of a fiber are derived as intensity-weighted averages over all the PPV points belonging to the fiber. Similarly, profiles of a physical quantity along the fibers can be calculated as the intensity-weighted averages over the PPV points with which the bone points are derived. Table \ref{tab:filament} lists the geometric and physical parameters of the fibers.   

To estimate the total gas masses of the fibers we need a conversion factor from $N_{\rm trc}$ to $N_{\rm H_2}$, i.e. the abundance. This is done by comparing $N_{\rm trc}$ obtained from the spectral fitting with $N_{\rm H_2}$ derived from the dust continuum emissions. We use the \moleA\ data to derive the total mass since its emission matches the $N_{\rm H_2}$ map the best among the three species (see Figure \ref{fig:m0}). A map of the \moleA\ column density is generated with the PPV points and was smoothed and resampled to match the resolution and gridding of the $N_{\rm H_2}$ map. An abundance map is then derived by dividing the $N_{\rm H^{13}CO^+}$ map with the $N_{\rm H_2}$ map. The resultant mean abundance for \moleA\ in the DR21(OH) ridge is $1.72\times10^{-10}$ with a regional fluctuation of $\sim0.25$ dex. In addition, we find that there is no strong correlation between the \moleA\ abundance and $T_{\rm dust}$ or $N_{\rm H_2}$ (see Appendix \ref{app:mass} for detailed analyses), which indicates that the \moleA\ abundance does not vary violently with the differentiated chemical conditions (e.g. depletion onto dust grains and formation/destruction through chemical reactions) and that adopting a constant abundance value is sufficient for calculating the mass. By converting $N_{\rm H^{13}CO^+}$ to $N_{\rm H_2}$ with this abundance we derive that the total masses of fibers \fa, \fb, and \fc\ to be \massFa, \massFb, and \massFc\ \msun, respectively, and the line masses are \lineMassFa, \lineMassFb, and \lineMassFc\ \msunpc, respectively, which is 1--2 orders of magnitude higher than typical low-mass filaments \citep[e.g.][]{2020arXiv201000006P}. See also Table \ref{tab:filament} for more details on the physical properties of the fibers.

\subsection{Free-fall Velocity Profile seen in Fiber \fc} \label{subsec:fall} 

One of the most prominent features of the DR21(OH) ridge in the PPV space is the large velocity gradient in fiber \fc, which is clearly seen in the PV diagrams of the \moleA\ and \moleB\ emissions (Figure \ref{fig:ppv}). This velocity gradient is not constant throughout the fiber, but increases from north to south, i.e., toward the direction of DR21(OH). In addition, the FWHM velocity dispersion of \fc\ increases almost monotonously from 0.3 to 1.9 \kms\ from north to south as seen in the \moleA\ data. These suggest that \fc\ is very likely influenced by the gravitational potential of the very massive dense core and the gas within the fiber is falling towards DR21(OH).  \citet{2010A&A...520A..49S} made a similar statement based on the emission distribution revealed by their single-dish observations. Here the data with a much higher angular resolution allow us to further test this interpretation, and we fit the PV pattern with a free-fall model. The PPV points used in the fitting are highlighted in Figure \ref{fig:ppv}. Based on the morphologies and spatial configuration of \fc\ and DR21(OH), we assume that \fc\ is falling directly toward DR21(OH) with no transverse motions and that the free-fall velocity is zero at infinity. In the 3D space, the free-fall velocity $v_{\rm ff}$ as a function of the distance to the mass center $l$ is 

\begin{equation}\label{eq:fall}
v_{\rm ff}(l)=\sqrt{\frac{2GM_{\rm c}}{l}},
\end{equation}

\noindent where $M_{\rm c}$ is mass of the gravity center. Since we can only observe LoS velocity and the projected distance on the PoS Eq. \ref{eq:fall} can be rewritten in a ``projected'' form that implicitly contains the inclination angle:

\begin{equation}\label{eq:fall_proj}
v_{\rm LoS}(l_{\rm p})=\sqrt{\frac{2GM_{\rm c, p}}{l_{\rm p}}} + v_{\rm c, LoS},
\end{equation}

\noindent where $l_{\rm p}=l{\rm cos}\theta_{\rm inc}$ and $M_{\rm c, p}=M_{\rm c}{\rm cos}\theta_{\rm inc}{\rm sin^2}\theta_{\rm inc}$ are the ``projected'' distance and mass, respectively, $\theta_{\rm inc}$ is the inclination angle of \fc\ against the PoS, and $v_{\rm c, LoS}$ is the systematic LoS velocity of the gravity center. Since $\theta_{\rm inc}$ is unknown we use the above equation to fit the velocity profile. The PPV points used for the fitting and the results are shown in Figure~\ref{fig:ppv}(g), which yield $M_{\rm c, p}=\Mp\pm\MpErr$ \msun\ and $v_{\rm c, LoS}=\vp\pm\vpErr$ \kms. The $v_{\rm c, LoS}$ value is well consistent with the results derived from single-dish spectral line observations toward DR21(OH) \citep{1973ApJ...182L..65M,1993MNRAS.261..694C}. The value of $M_{\rm c, p}$ is compatible with the mass of DR21(OH) \citep[446--1048~\msun,][]{2007A&A...476.1243M,2019ApJS..241....1C}, but since it is only the lower limit of the mass of the gravity center, it does not rule out the possibility that other MDCs around DR21(OH) also contribute to the gravitational attraction. We stress that the observations and fitting results all support a scenario that the gas in f3 is falling toward DR21(OH), and the latter seems to be the main (but not necessary the only) source of gravitational attraction. The mass accretion rate of \fc\ can be estimated by multiplying the line mass and the infall velocity, which is taken from the free-fall model (4.9~\kms without accounting for the projection effect; see Figure~\ref{fig:ppv}(g)), and is $\gtrsim2.5\times10^{-3}$~\msun~yr$^{-1}$, indicating that a mass feeding through f3 is capable of significantly increasing the core mass within a free-fall time (typically $10^5$~yr for dense cores) and thus play an important role in the high-mass and cluster formation within and around DR21(OH). 

\subsection{Instability of the fibers and the star formation in the DR21(OH) ridge} \label{subsec:instability} 

In this section we study the instability of the fibers and its relation to the star formation in this region. There are two modes of instabilities for a self-gravitating and isothermal gas cylinder: radial instability and axial instability. For the former, the critical line mass of a cylinder over which gravity overcomes supports of turbulence and thermal pressure is given by 

\begin{equation}\label{eq:radial}
\lambda_{\rm cr,radial}=\frac{2\sigma^2}{G},
\end{equation}

\noindent where $\sigma$ is the sound speed for thermal support and velocity dispersion in case of turbulent support \citep{1964ApJ...140.1056O,2000MNRAS.311..105F}. The fibers mostly fall in a temperature range of 15 to 25~K (Figure~\ref{fig:m0}(c)), so they are mildly supersonic with the velocity dispersions about 2 times the sound speed, leading to turbulent support. We then derive the critical line masses of 157, 182, and 130 \msunpc\ for fibers \fa, \fb, and \fc, respectively, which are only 7.3\%, 23\%, and 25\% of their actual line masses (see Table \ref{tab:filament}). This indicates that the fibers are unstable against radial gravitational collapse with the turbulent (and thermal) support. We further derive the line mass and critical line mass profiles along the fibers derived with the statistics of the bone points, and find that the line masses are $\sim10$ times larger than the critical values for most parts of the fibers (Figure \ref{fig:line_mass}). To estimate the relative importance of magnetic fields in the radial instability, we derive the magnetic energy, $E_{\rm B}$, and the gravitational binding energy, $E_{\rm G}$, of the fibers:

\begin{equation}\label{eq:E_B}
E_{\rm B}=\frac{\pi}{4}d^2L\cdot\frac{\bar{B}}{2\mu_0}=\frac{\pi^3}{32\mu_0}\bar{B}_{PoS}^2d^2L,
\end{equation} 

\begin{equation}\label{eq:E_G}
E_{\rm G}=\frac{GM^2}{(d^2L)^{1/3}},
\end{equation}

\noindent where $\mu_0$ is the vacuum permeability, and $d$, $L$, $M$, $\bar{B}$, $\bar{B}_{PoS}$ are the FWHM width, length, mass, average magnetic field strength, and average projected magnetic field strength on the PoS of a fiber, respectively. For $\bar{B}_{PoS}$ we adopt the value in \citet{2013ApJ...772...69G} (0.62 mG) derived from the JCMT observations. The resultant gravitational binding energies are 7.0$\times10^{48}$, 9.5$\times10^{47}$, and 2.9$\times10^{47}$ erg, for fibers \fa, \fb, and \fc, respectively, and the magnetic energies are 2.9$\times10^{46}$, 2.5$\times10^{46}$, and 1.3$\times10^{46}$ erg, which are only 0.4\%, 2.6\%, and 4.5\% of the gravitational energies, respectively. 

The axial instability (also known as the ``sausage'' instability) of a gas cylinder describes its mass agglomeration along the axial due to the growth of unstable perturbations. For such an instability there exists a characteristic length scale, $L_{\rm max,axial}$, delineating the nearly uniform spacing between the fragments, as a result of the fastest growing unstable perturbation \citep{1987PThPh..77..635N,2010ApJ...719L.185J}. \citet{1987PThPh..77..635N} find that for a self-gravitating isothermal cylinder, $L_{\rm max,axial}\sim22H$, where $H=c_s(4\pi G\rho_{\rm c})^{-1/2}$ is the scale height, $\rho_{\rm c}$ is the density at the center of the cylinder, and that this length scale does not change with the presence of an axial magnetic field. With the analytical solution for hydrostatic equilibrium in a self-gravitating isothermal cylinder \citep{1964ApJ...140.1056O}, we derive a relation between the FWHM width $d$ and the scale height as $d=4.336H$, and find that $L_{\rm max,axial}=0.61$, 0.57, and 0.47 pc for \fa, \fb, and \fc, respectively (see also Table \ref{tab:filament}). These length scales are $\sim4$ times shorter than the fiber lengths and the fibers are thus axially unstable, which is consistent with the fragmented appearance of the fibers in Figures \ref{fig:m0} and \ref{fig:rgb} and the oscillating line masses in Figure \ref{fig:line_mass}. In addition, the average separations of the line mass peaks (i.e. fragments) in Figure \ref{fig:line_mass} are 0.54, 0.40, and 0.42 pc for \fa, \fb, and \fc, respectively, which are in good agreement with their $L_{\rm max,axial}$ considering the projection effect. 

The analyses and results above all suggest that with the great line masses, gravity is playing a predominant role in shaping the dynamics of the fibers, and star formation is expected to happen in the fibers given their radial and axial instabilities. We use the catalogs of class II methanol masers and ultra-compact \hii\ (UC\hii) regions collected by \citet{2019ApJS..241....1C}, as well as the MCDs with strong SiO emissions in \citet{2007A&A...476.1243M}, to probe high-mass star-forming activities in the DR21(OH) ridge. To trace low-mass star formation we use the protostar catalogs in \citet{2007MNRAS.374...29D} and \citet{2014AJ....148...11K} generated with the \emph{Spitzer} data. All these sources are shown in Figures \ref{fig:m0}--\ref{fig:line_mass}. It is clear that \fa\ is active in high-mass star formation with a chain of active MDCs and UC\hii\ regions \citep{2007MNRAS.374...54K,2019ApJS..241....1C}. And it is widely known that the DR21(OH) core, where the three fibers conjoin, is forming a cluster of high-mass stars \citep{2012ApJ...744...86Z,2013ApJ...772...69G}. The very active (high-mass) star-forming activities in the DR21(OH) ridge may be attributed to a global collapse of the HFS \citep[see also][]{2010A&A...520A..49S}. The morphology and kinematics of the three fibers identified here provide new insights into this scenario. The enhancement in $N_{\rm H_2}$ from north to south in the ridge (Figure \ref{fig:m0}(b)), the transverse velocity gradient of fibers, and the free-fall velocity profile seen in \fc\ are all consistent with a physical picture that the three fibers are colliding in the south and that high-mass star formation is enhanced in the colliding region. On the other hand, there is no significant star formation in the northern parts of \fb\ and \fc, despite that they are still at least one order-of-magnitude more massive than low-mass star-forming filaments \citep[e.g.][]{2020arXiv201000006P}. In light of the great masses of DR21(OH) and fiber \fa, and the free-fall signature seen in \fc, the disruptive tidal field in this region may act as a force against local gravity and inhibit the star formation \citep{2010PhDT.........1R,2016arXiv160305720L}.

\subsection{Fibers of the DR21(OH) ridge in a bigger picture: relation with the outer filaments} \label{subsec:bigger} 

Besides the DR21(OH) ridge (or the DR21 ridge which includes the DR21 clump), another prominent feature of this massive HFS seen in the \emph{Herschel} maps is the lower-density filaments connected to the ridge/hub. These filaments stretch in all directions, and are mostly gravitationally unstable to form fragments \citep[i.e., filaments N, F1N, F1S, F3N, F3S, SW, and S in][]{2012A&A...543L...3H}, in contrast to the striations connected to low-mass filaments \citep{2008ApJ...680..428G,2013A&A...550A..38P}. Kinematic analyses using spectral observations show that these filaments are involved in the global collapse of the HFS and are probably feeding materials to the central regions \citep{2010A&A...520A..49S}. In comparison, these filaments have lengths comparable to the fibers seen in the ridge, but their line masses are a factor of a few to 10 lower \citet{2012A&A...543L...3H}; only the most massive filaments have masses and line masses comparable to those of the least massive fiber (\fc). 

One of the key questions regarding the outer filaments and inner fibers in this remarkable HFS is that whether the filaments are natural extensions of the fibers, or they are independent structures without actual connection. In Figure~\ref{fig:large} we plot the bones of our fibers and of the seven filaments in \citet{2012A&A...543L...3H} overlaid on the $N_{\rm H_2}$ map of \citet{2019ApJS..241....1C}. Apparently there are five filaments (N, F1N, F1S, F3N, and F3S) having potential connection with the fibers. Filament N seems to be the extension of fiber \fa\ in the map, yet there is possibility that it is instead the extension of fiber \fb. Filaments F1N/F1S are more likely to be the extension of fiber \fb, and single-dish $\rm ^{13}CO$ (1--0) observations in \citet{2010A&A...520A..49S} (see their Fig. 3) show that F1N/F1S joins the ridge at velocities of $-2$ to $-3$ \kms, well matching the velocity of \fb, further supporting that F1N/F1S is connected to \fb. \citet{2010A&A...520A..49S} suggested that F3 (resolved into F3N and F3S further out in Figure~\ref{fig:large}) is connected to a ``sub-filament'' seen in \lineA\ (corresponding to our fiber \fc). But F3N/F3S has a nearly east-west orientation whereas f3 is roughly in a north-south direction. There is also a velocity difference of 1--2 \kms\ between the two structures. Moreover, with a clear identification of the fibers based on our high resolution observations and an improved picture of the outer filaments constructed with the \emph{Herschel} data, the bones of the two structures do not connect with each other (Figure~\ref{fig:large}). Thus, it is more likely that f3 is an independent structure rather than being the inner part of F3.

Overall it appears that the filaments more parallel to the ridge (such as filaments N, F1N/F1S in Figure~\ref{fig:large}) are more likely to be the extensions of the inner fibers. This is consistent with a scenario that the large-scale filaments collide to form a ridge/hub, which will further grow in mass by gravitationally attracting the gas along the filaments, making the fibers in the ridge/hub more massive than the outer filaments. As the ridge/hub continues to gain mass, its gravitational well gets deeper and may attract new accretion flows with an even perpendicular orientation (such as filaments F3N/F3S in Figure~\ref{fig:large}). Future high resolution observations of massive HFSs covering both the hub/ridge and outer filaments may help to testify the above scenario and to advance our understanding of the initial conditions of high-mass star and cluster formation.

\section{Summary} \label{sec:conc}
We have observed the prominent DR21(OH) ridge in high-density tracing molecular spectral lines with the CARMA, and analyzed the data to extract the multiple velocity components, identified the fibers, and characterized the fiber properties, aimed at understanding the formation and evolution of the ridge and its relation to high-mass star and cluster formation. Our main findings are summarized below:

\begin{enumerate}

\item We clearly resolve the ridge into three fibers (\fa, \fb, and \fc) which are not seen in existing (sub)millimeter continuum observations. The fibers have lengths around 2~pc, widths about 0.1~pc, and are mildly supersonic with velocity dispersions about 2 times the sound speed. They are nearly parallel in the north and conjoin in the south around the most massive and active star-forming core DR21(OH).

\item A velocity gradient increasing from north to south is clearly seen in \fc, confirming previous single-dish observations \citep{2010A&A...520A..49S}. We find that the velocity gradient can be well reproduced with a model of gas flow in free-fall and that the infall rate is $\gtrsim2.5\times10^{-3}$~\msun~yr$^{-1}$. Thus the materials in \fc\ seem to be falling toward DR21(OH) due to strong gravitational attraction from the MDC (and possibly other MDCs around as well), a scenario also proposed by \citet{2010A&A...520A..49S}.

\item The three fibers have line masses significantly larger than the critical value, and from the instability analyses, they are unstable against gravitational collapse both axially and radially. While the most massive fiber, \fa, exhibits active high-mass star formation all the way from north to south, the other two fibers, \fb\ and \fc, show essentially no star formation in their northern parts. Given the free-fall velocity field seen in fiber \fc, gravitational tidal forces exerted by the massive materials in this region may account for some local inefficiency in star formation. 

\item On larger scales, some of the peripheral filaments are likely to be the outer extensions of the fibers inside the ridge. By comparing the morphologies and masses between the peripheral filaments and the ridge fibers, we speculate that large filaments of more parallel orientations collide to form a HFS, and the central hub/ridge continues to grow in mass by gravitationally attracting the gas from the outer parts of the filaments, feeding the central colliding parts into massive fibers and promoting high-mass star and cluster formation. As the ridge gets more massive, its strong gravitational attraction may induce additional accretion flows of all the orientations.

\end{enumerate}

\acknowledgments 
Y.C. and K.Q. are partially supported by National Key R\&D Program of China No. 2017YFA0402600, and acknowledge the support from National Natural Science Foundation of China (NSFC) through grants U1731237 and 11629302. Y.C. acknowledges the funding supports from the Scholarship No. 201906190105 of the China Scholarship Council and from the Predoctoral Program of the Smithsonian Astrophysical Observatory (SAO). G.L. acknowledges the supports from Yunnan University grant C176220100028. 

\facilities{Telescope}

\software{Astropy\citep{2013A&A...558A..33A}, SciPy\citep{scipy}, scikit-learn\citep{scikit-learn}, MIRIAD\citep{1995ASPC...77..433S}} 

\clearpage

\startlongtable
\begin{deluxetable}{l|ccc}
\tabletypesize{\small}
\tablecaption{Properties of the fibers\label{tab:filament}}
\tablehead{\colhead{Property}& \colhead{\fa}& \colhead{\fb}& \colhead{\fc}}
\startdata
\moleA\ detection&  Full&   Full&   Full\\
\moleB\ detection&  Full&   Full&   Southern part\\
\moleC\ detection&  Full&   Northern part&   None\\
\hline
Length (pc)&                  2.40& 2.31&   1.80\\
Width (pc)&                   0.12& 0.11&   0.09\\
${\rm v}_{\rm lsr}$ (\kms)&         $-4.22$&    $-2.28$&  $-0.84$\\
$\sigma_{\rm v}$ (\kms)&  0.58& 0.62&   0.53\\
Mass (\msun)&                 5117&   1838& 921\\
$\lambda$ (\msunpc)&          2133&   795&  511\\
$\lambda_{\rm cr,radial}$ (\msunpc)\tablenotemark{a}& 157&    182&  130\\
$L_{\rm max,axial}$ (pc)\tablenotemark{b}& 0.61&  0.57&   0.47\\
\enddata
\tablenotetext{a}{Critical line mass for radial instability.}
\tablenotetext{b}{Characteristic length scale  for axial instability, also the wavelength of fastest growing perturbations.}
\end{deluxetable}

\begin{figure*}[htb!]
\epsscale{1.}\plotone{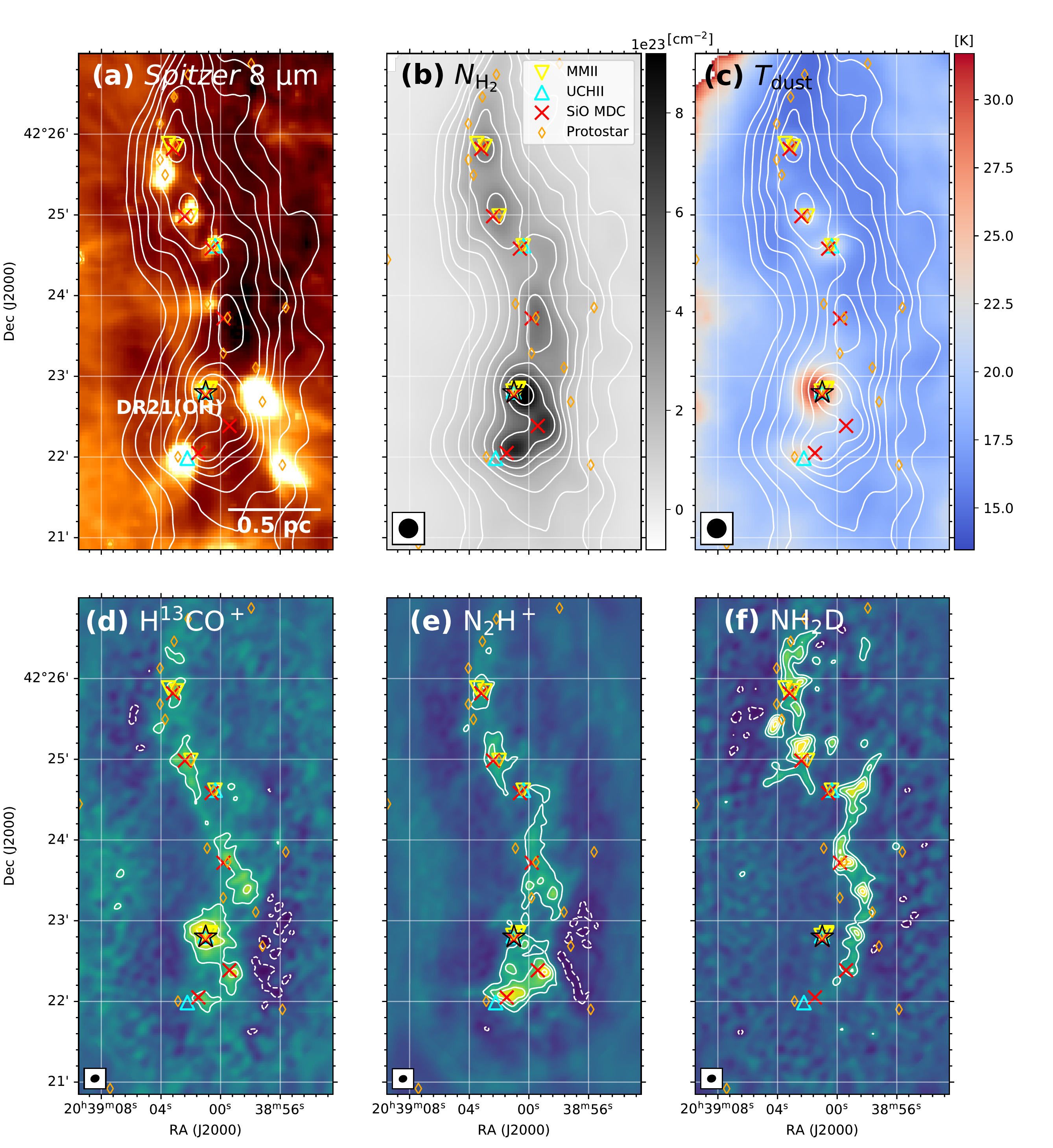}
\caption{\emph{Spitzer} 8 \um\ map (Panel a), high-resolution maps of $N_{\rm H_2}$ (Panel b) and $T_{\rm dust}$ (Panel c), and moment-0 maps of the three molecular transitions (Panels d--f) of the DR21(OH) ridge. White contours of $N_{\rm H_2}$ in Panels a--c are drawn with the same levels as those in Figure \ref{fig:channel}. Contours of the moment-0 maps in Panels d--f are drawn in $[...,-3,-2,-1,1,2,3,...]\times \delta I$, where $\delta I=$1, 8, and 1 $\rm Jy\ beam^{-1}km\ s^{-1}$ for \moleA, \moleB, and \moleC, respectively, with negative contours shown as dashed curves. Other symbols in all the panels are the same as those in Figure \ref{fig:channel}.}\label{fig:m0}
\end{figure*}

\begin{figure*}[htb!]
\epsscale{1.2}\plotone{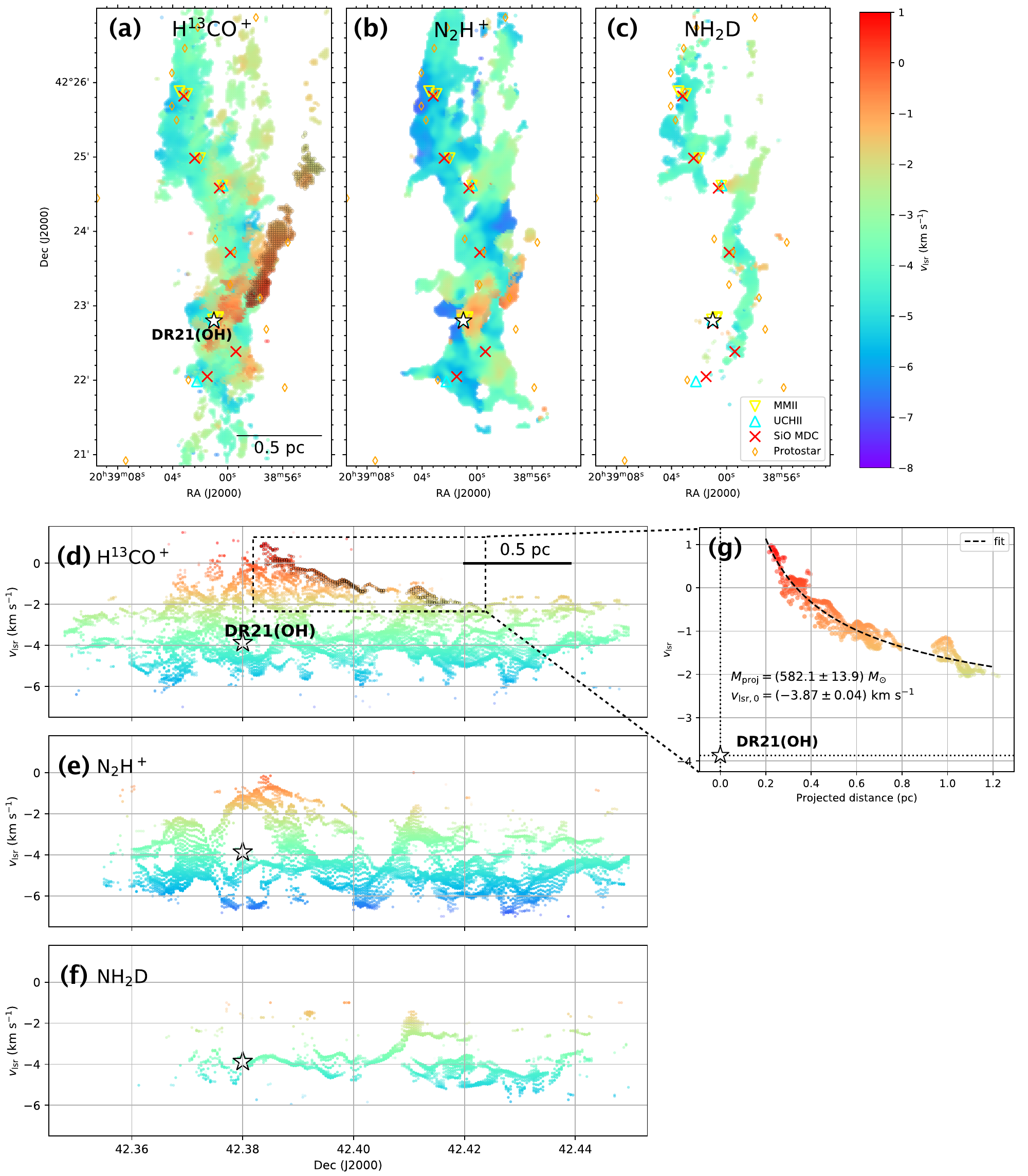}
\caption{2D views of the fitted velocity components (PPV points, see Sect. \ref{subsec:fit}), which are shown in colored dots with the velocity color coding following a color bar to the top right. (a--c) PPV points (derived with the \moleA, \moleB, and \moleC\ spectra, respectively) placed on the plane of sky. (d--f) Position-velocity (PV) plots in the \moleA, \moleB, and \moleC\ lines, respectively, of the PPV points. Dots highlighted by black circles in Panels (a) and (d) denote the PPV points that are used for fitting a free-fall model shown as a dashed curve in Panel (g) (see Sect. \ref{subsec:fall} for details). Other symbols in all the panels are the same as shown in Figure \ref{fig:channel}.}\label{fig:ppv}
\end{figure*}

\begin{figure*}[htb!]
\epsscale{1.2}\plotone{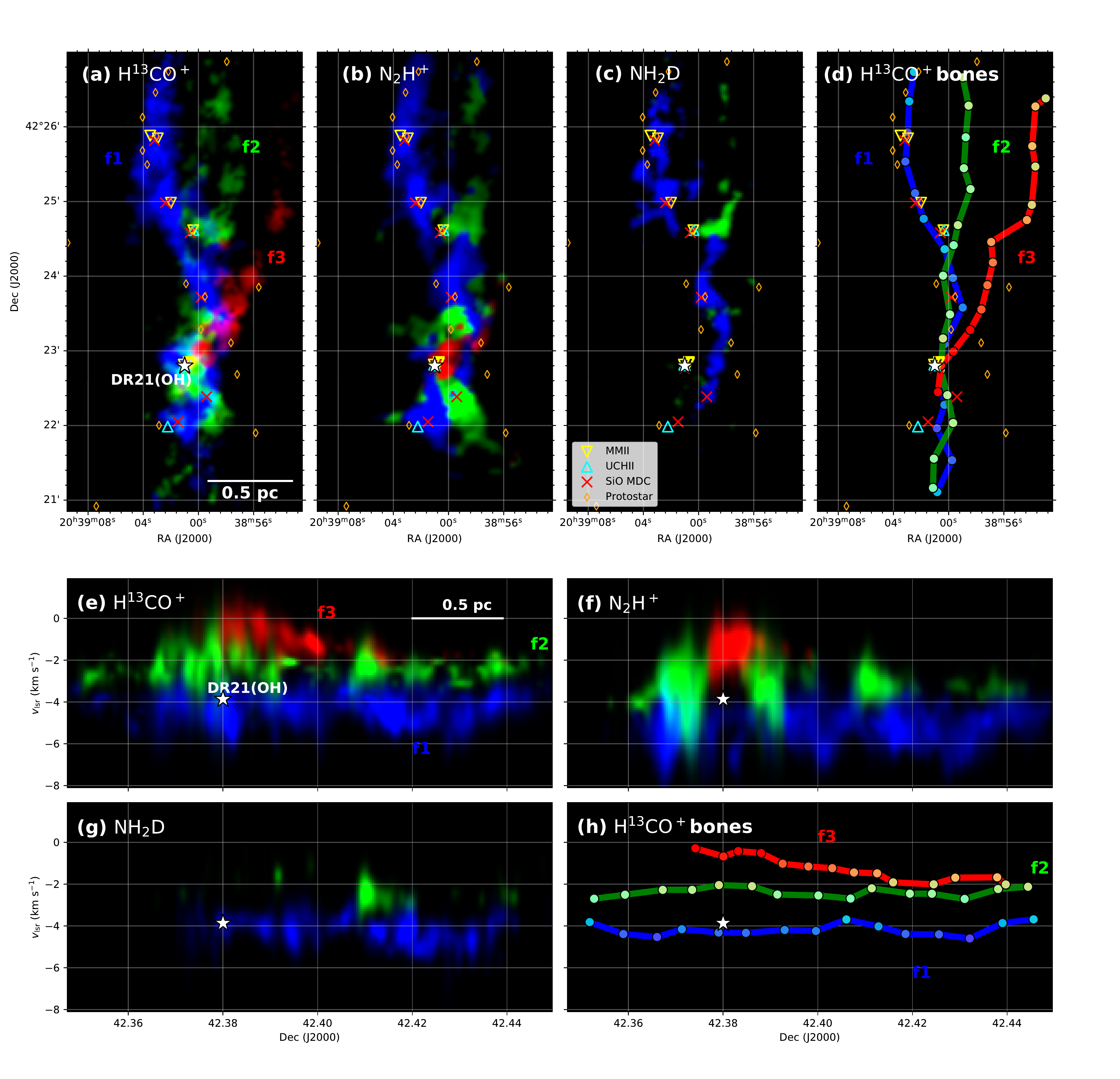}
\caption{Panels (a--c) show the fibers, identified with the \moleA, \moleB, and \moleC\ spectra, respectively, as seen projected on the PoS; Panels (e--f) show the identified fibers in the \moleA, \moleB, and \moleC\ PV diagrams, respectively. The three fibers, f1, f2, and f3, are coded to blue, green, and red, respectively in all the panels, and the brightness in Panels (a--c) and (e--g) are proportional to the intensity that is derived from the spectral line fitting (Sect. \ref{subsec:fit}). The bones of the fibers identified with the \moleA\ data are shown in Panels (d) and (h). The overlapping of the three bones in Panel (d) reflects their presumed positions along LoS, i.e. \fa, \fb, and \fc\ from far side to near side (Sect. \ref{subsec:iden}). Other symbols in all the panels are the same as those in Figure \ref{fig:channel}.}\label{fig:rgb}
\end{figure*}

\begin{figure*}[htb!]
\epsscale{.8}\plotone{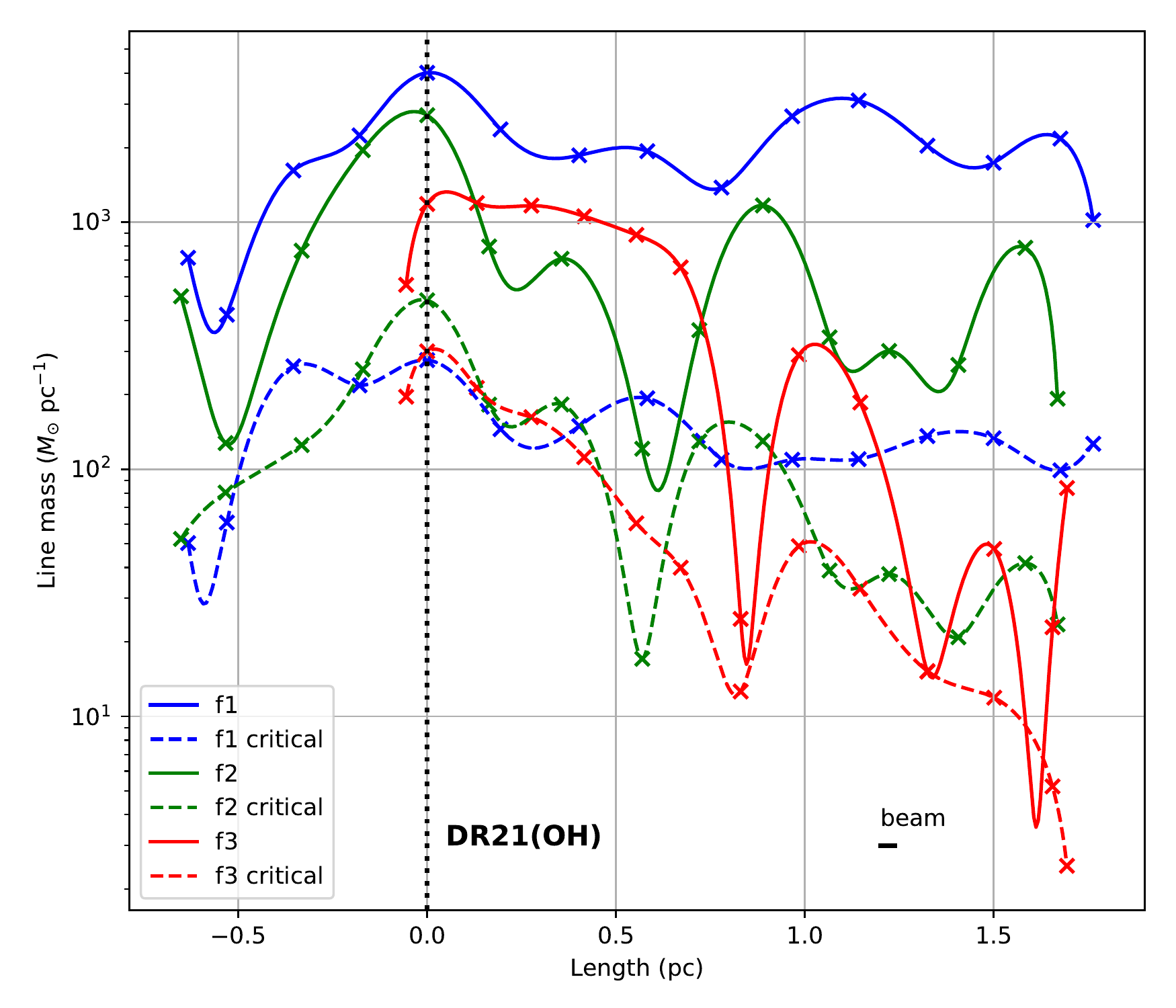}
\caption{Line mass (solid lines) and critical line mass (dashed lines) profiles along the bones of the three fibers, derived with the \moleA\ data. The data points are marked as crosses and the lines are interpolations with order-3 splines. Critical line masses are derived with Eq. \ref{eq:radial} and reflect the radial instability of the fibers (Sect. \ref{subsec:instability}). All the profiles are aligned in position such that the origins are the closest bone points to DR21(OH). The CARMA beam is shown as a dash.}\label{fig:line_mass}
\end{figure*}

\begin{figure*}[htb!]
\epsscale{1.}\plotone{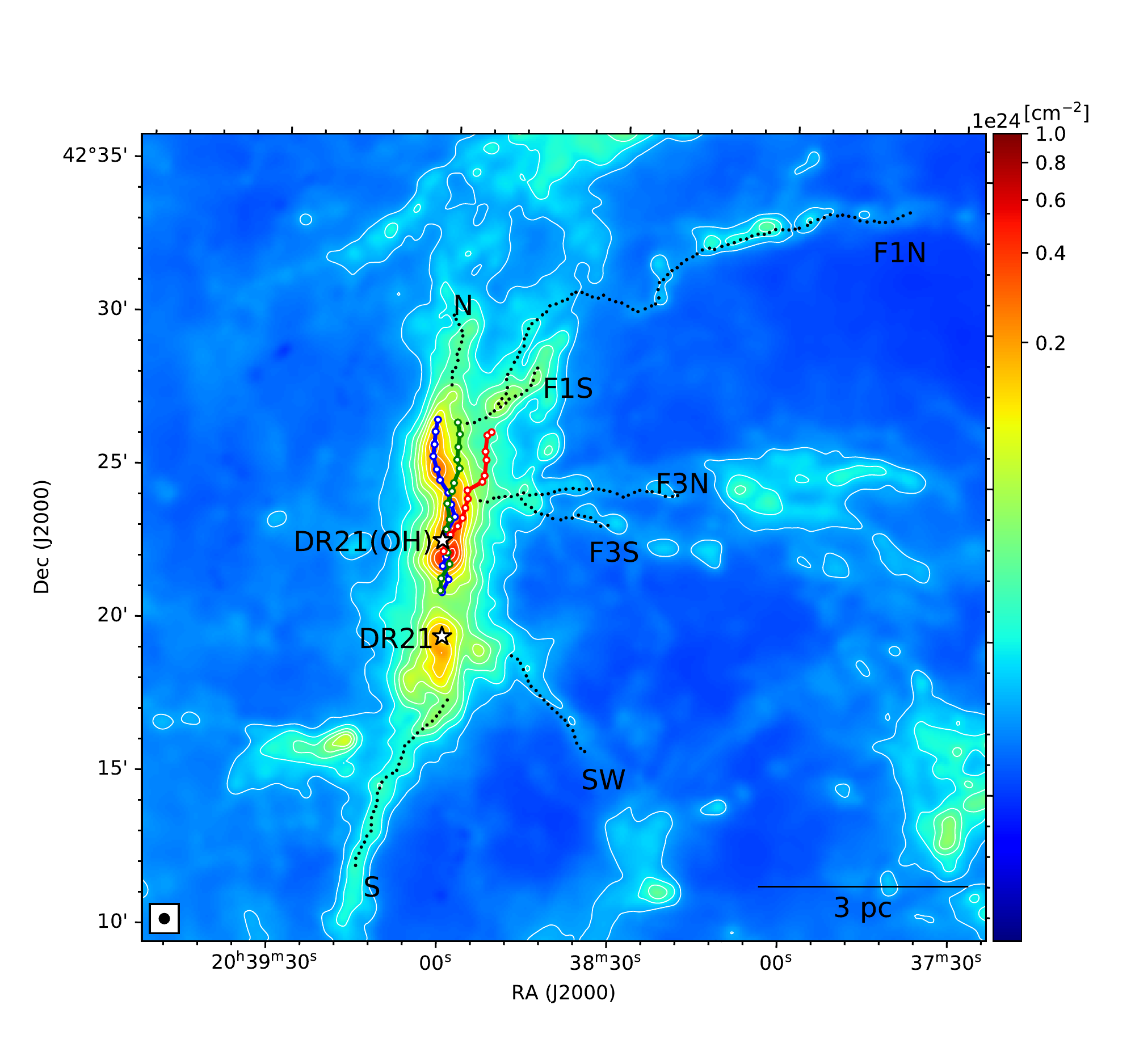}
\caption{The $N_{\rm H_2}$ map derived by \citet{2019ApJS..241....1C} showing the DR21(OH) ridge and its large-scale environment. Contours of the $N_{\rm H_2}$ are drawn in $1.5^{[0,1,2,3,...]}\times1.2\times10^{22}\rm\ cm^{-2}$. The bones of the three fibers in the DR21(OH) ridge are shown as blue, green, and red lines. Dotted lines outline the peripheral filaments in this region identified by \citet{2012A&A...543L...3H}. The two most well-known MDCs, DR21 and DR21(OH), are marked as stars. Beam size of the map (18\arcsec.4) is shown in the lower-left corner.}\label{fig:large}
\end{figure*}

\clearpage


\bibliography{ms}{}
\bibliographystyle{aasjournal}

\clearpage
\appendix 


\counterwithin{figure}{section}

\section{Robustness of the $N_{\rm H_2}$ and $T_{\rm dust}$ maps}\label{app:6band}

\begin{figure*}[htb!]
\epsscale{1}\plotone{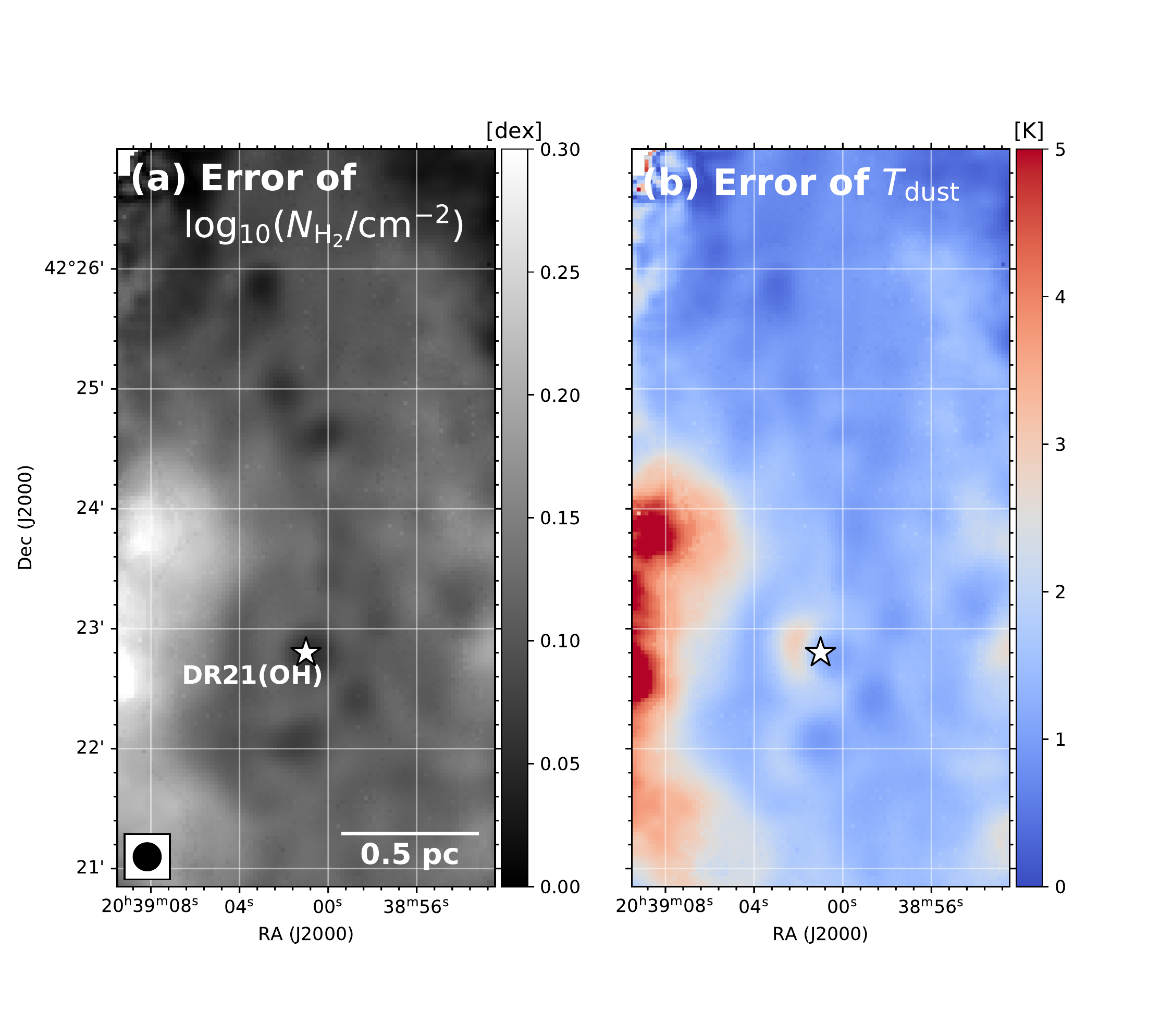}
\caption{Maps of the 1-$\sigma$ uncertainties of (a) $N_{\rm H_2}$ and (b) $T_{\rm dust}$ derived from the SED fitting procedure (Sect. \ref{subsec:N_map}).}\label{fig:err_map}
\end{figure*}

In this section we discuss the uncertainties in the $N_{\rm H_2}$ map and the $T_{\rm dust}$ map (Figure \ref{fig:m0}) of the DR21(OH) ridge. The maps are derived through fitting the spectral energy distributions (SEDs) obtained from the continuum images in the three bands of \emph{Herschel} 160 \um, JCMT 450 and 850 \um\ with a modified blackbody model (see Sect. \ref{subsec:N_map}). The flux uncertainty of each band used in the fitting is estimated as 20\%. The 1-$\sigma$ uncertainty maps of ${\rm log_{10}}N_{\rm H_2}$ and $T_{\rm dust}$ derived in the SED fitting are shown in Figure \ref{fig:err_map}, which give an mean value of 0.11 dex and 1.6 K, respectively. 

In the reduction of JCMT/SCUBA-2 data a largest angular scale is used to filter the independent low-frequency noise on the images \citep[see][]{2015MNRAS.454.2557M}. Its effect on the SED fitting can be seen as rises of temperature and temperature uncertainty on the map edges (Figures \ref{fig:m0}b and \ref{fig:err_map}b). For both our 450 and 850 \um\ data this scale is 480\arcsec\ (3.5 pc@1.5 kpc), which is larger than our map sizes and thus has limited filtering effects in our region of interest (the three fibers in the middle of the maps).

The 850 \um\ continuum flux at DR21(OH) could potentially be contaminated by the strong free-free emission of this region. To evaluate this effect we adopt the radio SED relation of DR21(OH) in \citet{2009ApJ...698.1321A} (see their Figure 5) and extrapolate it to 850 \um, which yields 66 mJy per \beamColden\ beam. This is $\sim$1/400 of the actual continuum intensity of DR21(OH) at 850 \um\ and can be neglected.

\begin{figure*}[htb!]
\epsscale{1}\plotone{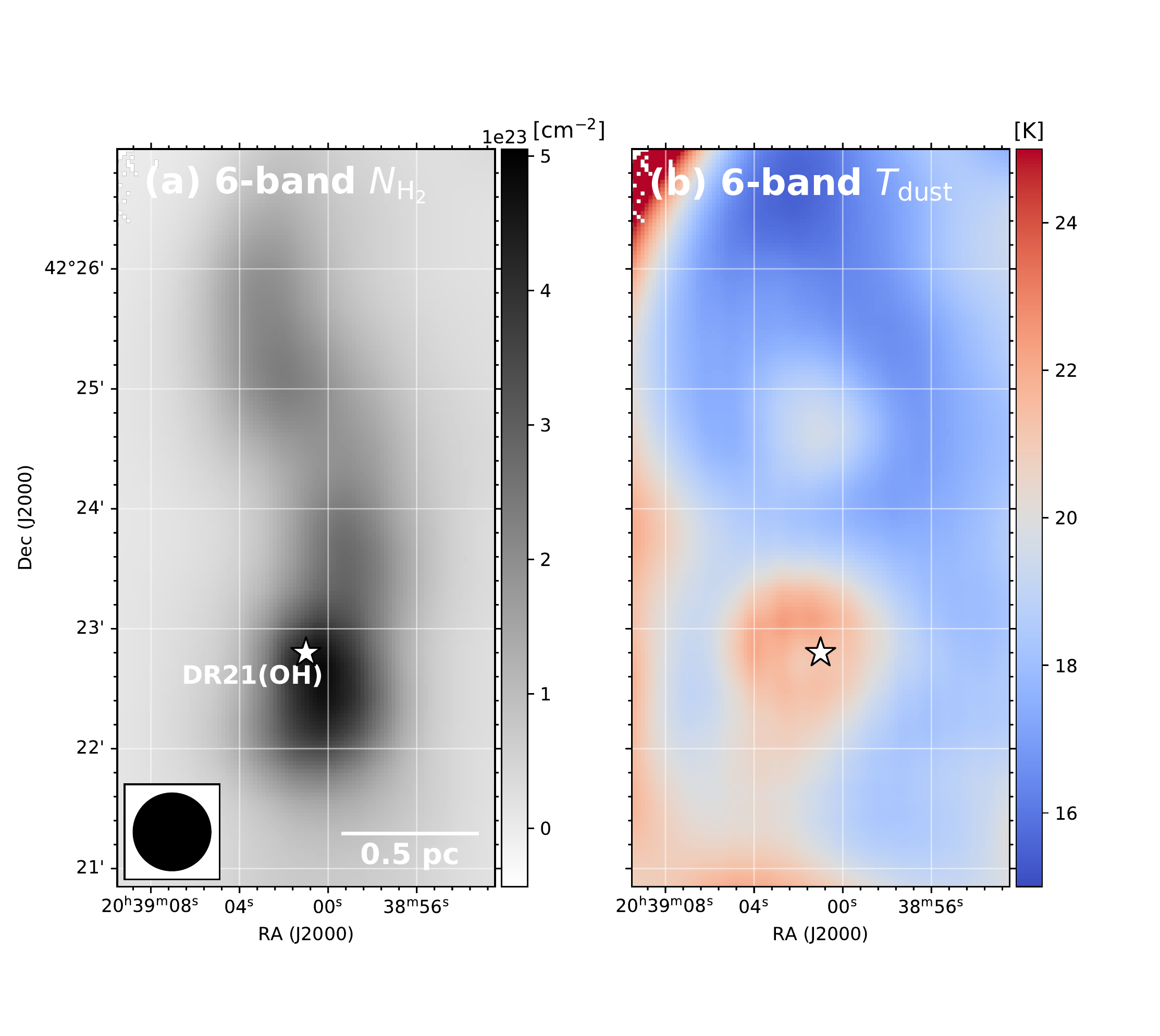}
\caption{Maps of (a) $N_{\rm H_2}$ and (b) $T_{\rm dust}$ derived with the 6-band continuum data (\emph{Herschel} 160, 250, 350, and 500 \um; JCMT 450 and 850 \um).}\label{fig:NT6}
\end{figure*}

\begin{figure*}[htb!]
\epsscale{1}\plotone{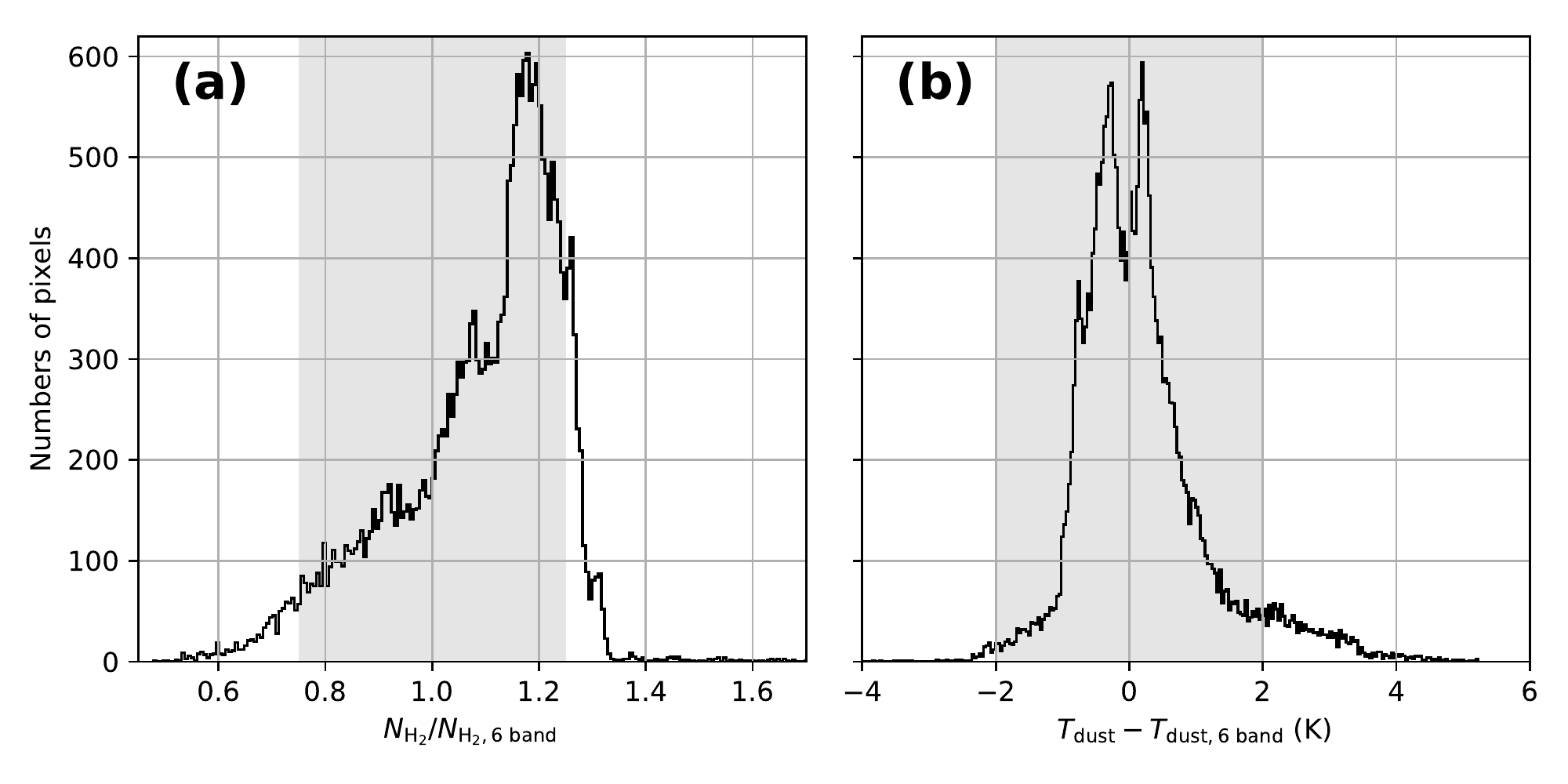}
\caption{Distributions of the differences in (a) $N_{\rm H_2}$ and (b) $T_{\rm dust}$ between the 3-band data and the 6-band data. The gray shaded regions show the pixels with (a) $|N_{\rm H_2}/N_{\rm H_2,6\ band}|\le$25\% and (b) $|T_{\rm dust}-T_{\rm dust,6\ band}|\le$2 K, which take up 90\% and 93\% of the total number of pixels, respectively.}\label{fig:hist_err6}
\end{figure*}

To examine the robustness of the 3-band SED fitting we derive another set of $N_{\rm H_2}$ and $T_{\rm dust}$ maps using the same techniques in deriving the 3-band ones but with continuum images in 6 bands of \emph{Herschel} 160, 250, 350, 500 \um, and JCMT 450 \& 850 \um. The resultant maps have a coarser resolution of \beamColdenSix\ and are presented in Figure \ref{fig:NT6}. The 3-band maps were then smoothed to this resolution and compared with the 6-band values as in Figure \ref{fig:hist_err6}. As indicated in the figure, 93\% of the pixels in the temperature map have differences less than 2 K, and 90\% of the pixels in the column density map have relative differences less than 25\%. The $N_{\rm H_2}$ values derived with the 3-band data are on average 9\% higher than the 6-band results, which is a result of the higher \emph{Herschel} intensities at longer wavelengths compared with the JCMT ones (due to the large-scale filtering effect of the latter). The two fitting results agree with each other, with the overall uncertainties of fiber masses (see Appendix \ref{app:mass}).

\section{Uncertainties of the fiber masses}\label{app:mass}

\begin{figure*}[htb!]
\epsscale{1}\plotone{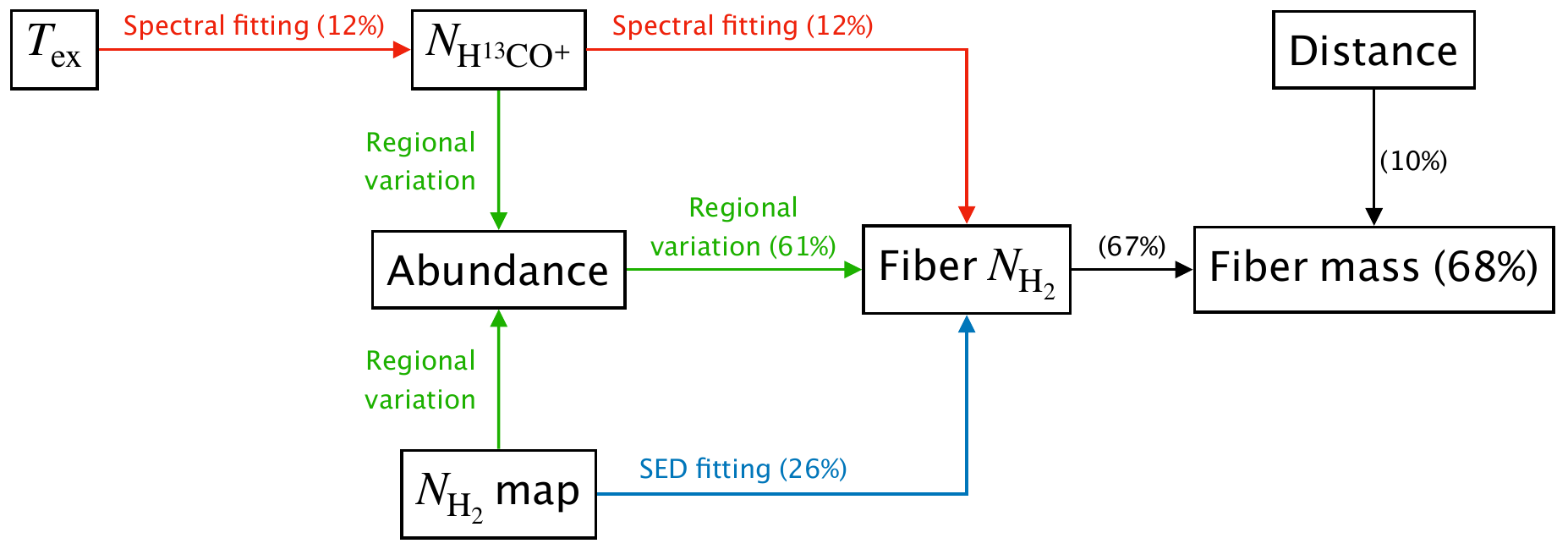}
\caption{Diagram showing the sources of errors in fiber masses and their transfer along the derivation procedure. The red, green, and blue connecting arrows represent the three major sources of errors in $N_{\rm H_2}$ of fibers discussed in Sect \ref{app:mass}. Percentages in brackets on the connecting arrows are the uncertainties of the \emph{next} blocks contributed from the previous ones. The overall uncertainty of fiber masses is estimated to be 68\%.}\label{fig:diagram}
\end{figure*}

In this section we estimate the uncertainties of fiber masses. Figure \ref{fig:diagram} presents the main sources of uncertainties in the derivation procedure of fiber masses, which can be divided into (1) uncertainties of assuming $T_{\rm ex}=T_{\rm dust}$ in fitting the \moleA\ spectra; (2) regional variations of the \moleA\ abundance; (3) SED fitting errors of the $N_{\rm H_2}$ as discussed in Sect. \ref{app:6band}; (4) distance uncertainty in parallax measurement. Here we discuss them respectively. 

\begin{figure*}[htb!]
\epsscale{1}\plotone{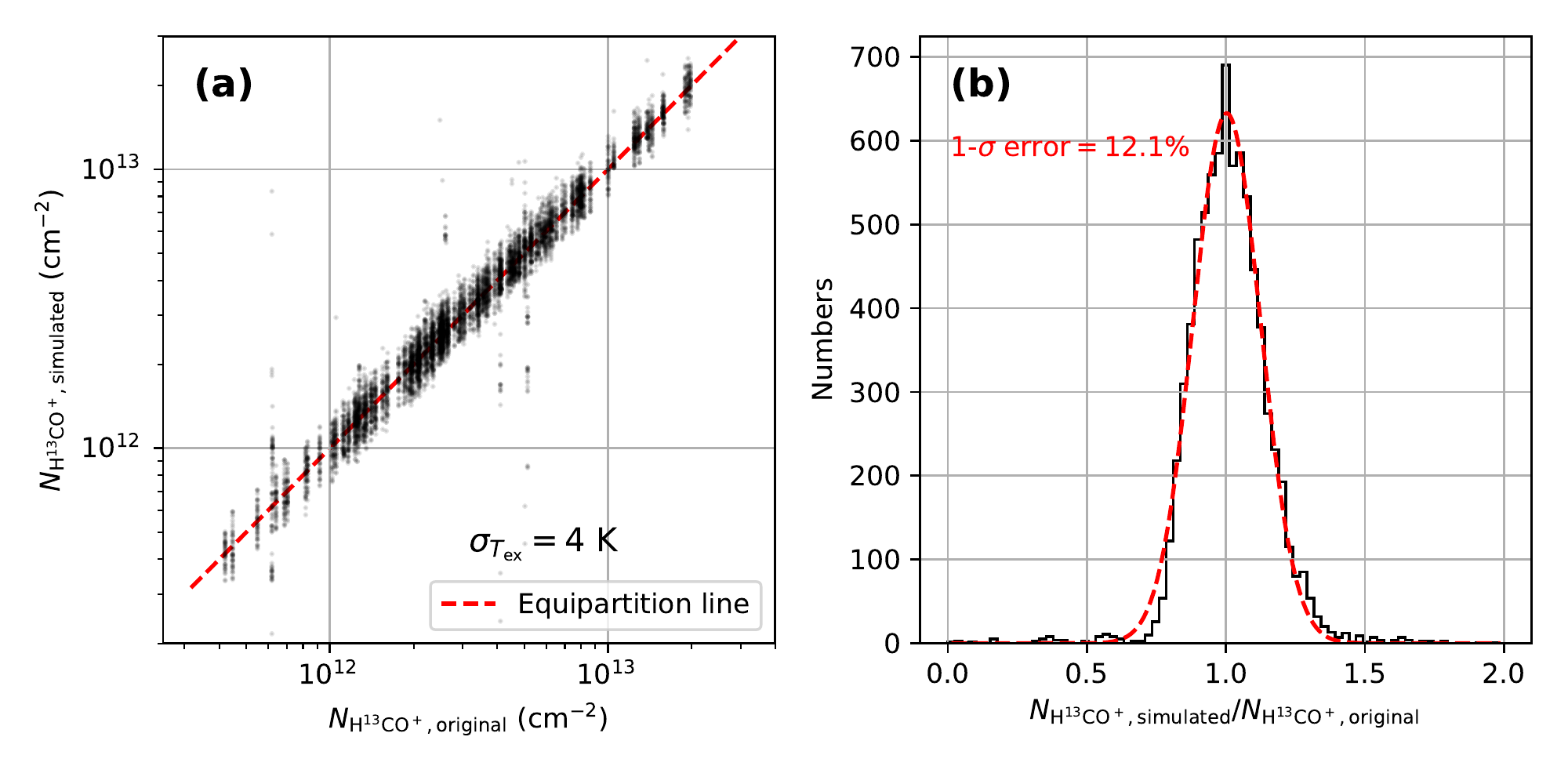}
\caption{(a) \moleA\ column densities derived from the original spectral fitting (Sect. \ref{subsec:fit}) versus those derived with simulated excitation temperatures (Appendix \ref{app:mass}). (b) Distributions of the ratios of the two column densities. A Gaussian fitting is shown as the red dashed line.}\label{fig:spectral_mc}
\end{figure*}

In the spectral fitting procedure we made the assumption of $T_{\rm ex}=T_{\rm dust}$ (Sect. \ref{subsec:fit}). This assumption seems to be reasonable for the DR21(OH) ridge since the $\rm H_2$ volume densities of its most parts exceed a few $10^4\rm\ cm^{-3}$, which is sufficient for the dust and gas to be thermally coupled via collision \citep{1991ApJ...377..192H}. At positions where the fibers are seen, the density is even higher. Here we implement the spectral fitting with varying $T_{\rm ex}$ to study its influence on the resultant tracer density $N_{\rm \rm H^{13}CO^+}$. We randomly select 1,000 out of the 7,625 \moleA\ spectra and for each spectrum we assign the fixed $T_{\rm ex}$ parameter with 50 random values respectively and do the spectral fitting repeatedly. The random temperature values observe a normal distribution with a mean equal to the dust temperature of that pixel and a standard deviation of 4 K. The resultant $N_{\rm \rm H^{13}CO^+}$ values obtained from the spectral fitting are compared with the original ones. As shown in Figure \ref{fig:spectral_mc}, the uncertainties in $T_{\rm ex}$ introduce a 1-$\sigma$ relative error of 12.1\% in $N_{\rm \rm H^{13}CO^+}$ and $N_{\rm H_2}$ of the fibers. 

\begin{figure*}[htb!]
\epsscale{1}\plotone{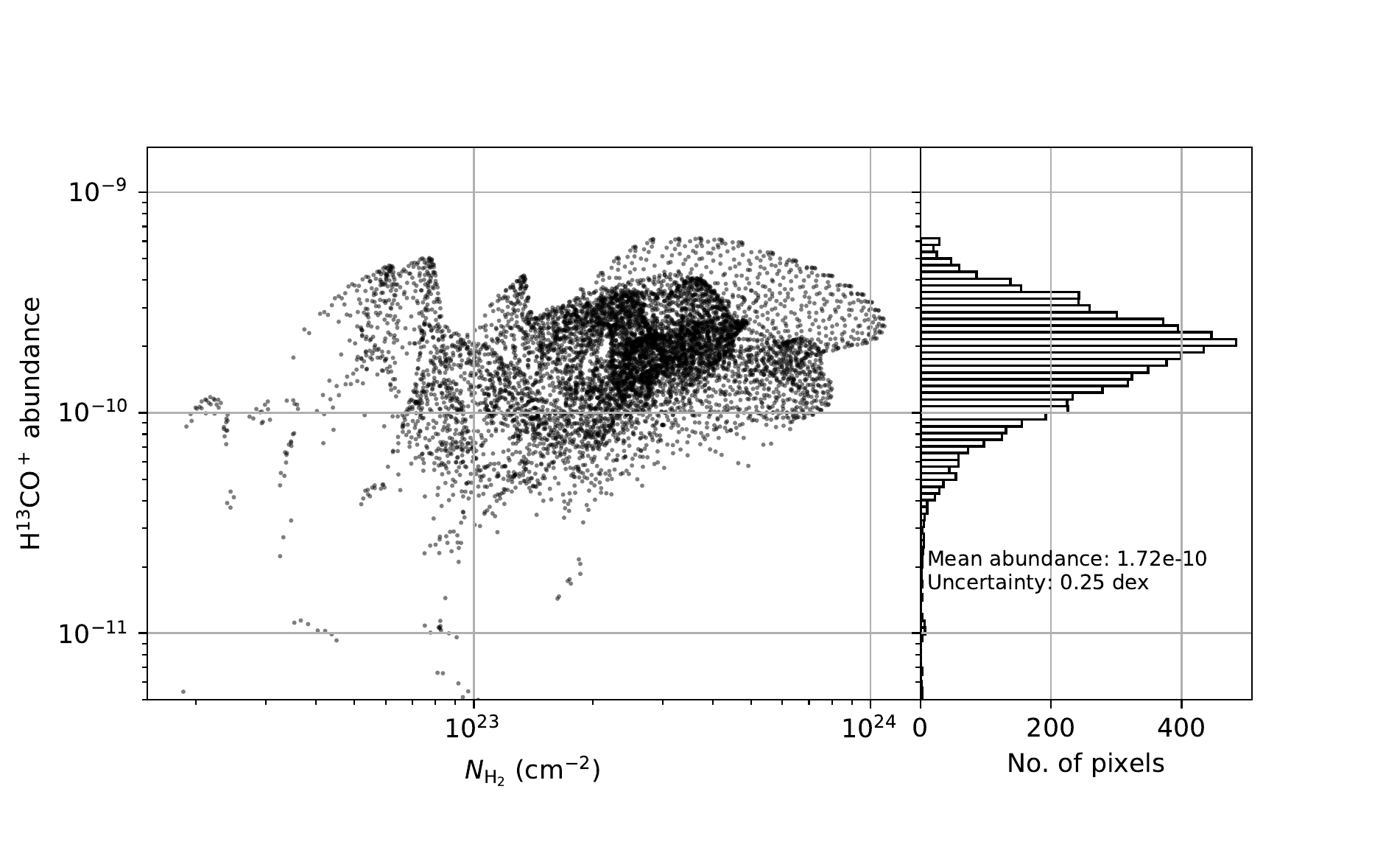}
\caption{$\rm H_2$ column density versus the abundance of \moleA\ of each pixel of the $\rm H_2$ column density map (Figure \ref{fig:m0}b). Histogram of the \moleA\ abundance is also shown on the right.}\label{fig:abundance}
\end{figure*}

When converting $N_{\rm \rm H^{13}CO^+}$ to $N_{\rm H_2}$ for the fibers a constant \moleA\ abundance is assumed (Sect. \ref{subsec:char}), which can be violated if the abundance has regional variations. To compare the two column densities we generated a $N_{\rm \rm H^{13}CO^+}$ map with the spectral fitting results, smoothed it to the resolution of the $N_{\rm H_2}$ map (Figure \ref{fig:m0}b), and re-gridded it to the same pixel frame. Figure \ref{fig:abundance} plots the relation of $N_{\rm H_2}$ versus the abundance and the distribution of the abundance. There is no significant correlation between the two quantities, which indicates that $N_{\rm \rm H^{13}CO^+}$ is an overall good tracer of $N_{\rm H_2}$ and that adopting a constant abundance value is robust. The 1-$\sigma$ regional variation of the \moleA\ abundance is 0.25 dex, which results in a relative uncertainty of 61\% in $N_{\rm H_2}$ of the fibers.    

To summarize, the uncertainty of $T_{\rm ex}$, the regional variation of \moleA\ abundances, and the errors in SED fitting contribute independently to the relative uncertainty of fiber $N_{\rm H_2}$ by 12.1\%, 61\%, and 25.6\%, respectively, which yield an overall uncertainty of 67.3\%. For fiber masses, the uncertainty in distance measurements should be considered given $M\propto N_{\rm H_2}D^2$. We adopt the number of 5\% in \citet{2012A&A...539A..79R}, which causes an uncertainty of 10\% in mass. The overall uncertainty of fiber masses is then estimated to be 68\%.


\end{document}